\documentclass[acmsmall,nonacm]{acmart} 
\AtBeginDocument{%
  }

\setcopyright{acmlicensed}
\copyrightyear{2018}
\acmYear{2018}
\acmDOI{XXXXXXX.XXXXXXX}
\acmConference[Conference acronym 'XX]{Make sure to enter the correct
  conference title from your rights confirmation email}{June 03--05,
  2018}{Woodstock, NY}
\acmISBN{978-1-4503-XXXX-X/2018/06}

\usepackage[T1]{fontenc}
\usepackage{graphicx}

\usepackage{verbatim}
\usepackage{enumitem}
\usepackage{csquotes}
\usepackage{alltt}
\usepackage{caption}
\usepackage{subcaption}
\usepackage{wrapfig}
\newcommand{\etal}{\textit{et al.}}
\usepackage{algorithm}
\usepackage{algorithmic}
\usepackage{tikz}
\usepackage{tikz-network}
\usetikzlibrary{positioning, quotes, arrows.meta}
\usepackage{amsmath}
\usepackage{mathpartir}
\usepackage{bussproofs}
\usepackage{pifont}
\usepackage{xcolor}
\usepackage{mathrsfs}
\usepackage{alltt}
\usepackage{pifont}

\usepackage{fontawesome5}

\newcommand{\singlebox}[1]{%
  \tikz[baseline={(node.base)}]{\node[draw, rounded corners, inner sep=1.5pt] (node) {\texttt{#1}};}%
}

\newcommand{\dashedgraysinglebox}[1]{%
  \tikz[baseline={(node.base)}]{%
    \node[
      dashed,
      draw=gray,
      text=gray,
      rounded corners,
      inner sep=2.0pt
    ] (node) {\texttt{#1}};
  }%
}

\newcommand{\doublebox}[1]{%
  \tikz[baseline={(node.base)}]{\node[double, draw, rounded corners, inner sep=1.5pt] (node) {\texttt{#1}};}%
}

\definecolor{lightpink}{RGB}{255, 200, 220}
\definecolor{lightyellow}{RGB}{255, 255, 200}

\begin{document}

\title{Abduction Prover in Isabelle/HOL}

\author{Yutaka Nagashima}
\orcid{0000-0001-6693-5325}
\affiliation{%
  \institution{Institute of Computer Science, the Czech Academy of Sciences}
  \city{Prague}
  \country{Czechia}
}

\author{Daniel Sebastian Goc}
\orcid{0000-0002-2347-8037}

\renewcommand{\shortauthors}{Nagashima et al.}

\begin{abstract}
Proof assistants based on expressive logics suffer limited automation for proof search, raising the cost of formal verification based on proof assistants.
We address this problem by introducing the Abduction Prover for Isabelle/HOL.
Given a challenging proof goal, the Abduction Prover constructs a proof script for the goal by identifying useful conjectures using abductive reasoning.
\end{abstract}





\settopmatter{printacmref=false}

\setcopyright{none}
\maketitle

\section{Introduction} \label{sec:intro}
Consider the following proof goal presented in a Haskell-like syntax.
\begin{alltt}
data List \(\alpha\) = [ ] | \(\alpha\) : (List \(\alpha\))
+ :: List \(\alpha\) => List \(\alpha\) => List \(\alpha\)
+   [ ]   \(ys\) = \(ys\)
+  (\(x\):\(xs\)) \(ys\) = \(x\):(\(xs\) +  \(ys\))

rev1 :: List \(\alpha\) => List \(\alpha\) => List \(\alpha\)
rev1  [ ]   ys = ys
rev1 (x:xs) ys = rev1 xs (x:ys)

rev2 :: List \(\alpha\) => List \(\alpha\)
rev2  [ ]   = [ ]
rev2 (\(x\):\(xs\)) = (rev2 \(xs\)) + (\(x\):[ ])

theorem revs_eq: "rev1 \(xs\) [ ] = rev2 \(xs\)"
\end{alltt}

This theorem, named \verb|revs_eq|, states the equivalence of two reverse functions, \verb|rev1| and \verb|rev2|, defined on lists of arbitrary types in Isabelle/HOL's underlying logic: classical higher-order logic on simply typed lambda calculus.
Conventionally, Isabelle users apply tools called proof tactics to transform proof goals to develop proof scripts.
Proof tactics are meta-programming tools that transform proof goals at hands into shapes closer to the obviously correct statement \verb|True|.
Isabelle offers more than 200 tactics, and it is users' responsibility to keep applying appropriate tactics to proof goals until they become obviously true.
This approach is called tactical theorem proving, and this is the dominant style adopted by the major interactive theorem provers (ITPs) such as Rocq \cite{coq}, Lean \cite{lean}, and the HOL prover \cite{hol4}.
For example, the first step to prove this goal is to apply the induction tactic as follows:

\begin{alltt}
theorem revs_eq: "rev1 \(xs\) [] = rev2 \(xs\)" apply (induct \(xs\))
\end{alltt}

This leads to two new sub-goals, corresponding to the base case and the step case of the structural induction on the variable \textit{xs}.

\begin{alltt}
subg1. rev1 [] [] = rev2 []
subg2. rev1 \(xs\) [] = rev2 xs \(\Longrightarrow\) rev1 (\(x1\):\(xs\)) [] = rev2 (\(x1\):\(xs\))
\end{alltt}

The second sub-goal itself is a significant challenge. 
Further applications of tactics to the second sub-goal may complicate it.
Addressing it effectively requires first conjecturing several auxiliary lemmas and proving them. Generally, when using ITPs, we have to address the following question:

\begin{enumerate}[label=\textsf{Q\arabic*:}]
    \item What tactics should we apply with which arguments?
    \item When should we stop applying tactics and introduce auxiliary lemmas?
    \item What auxiliary lemmas should we introduce?
\end{enumerate}

\begin{figure}
    \centering
    \includegraphics[
        width=\textwidth
    ]{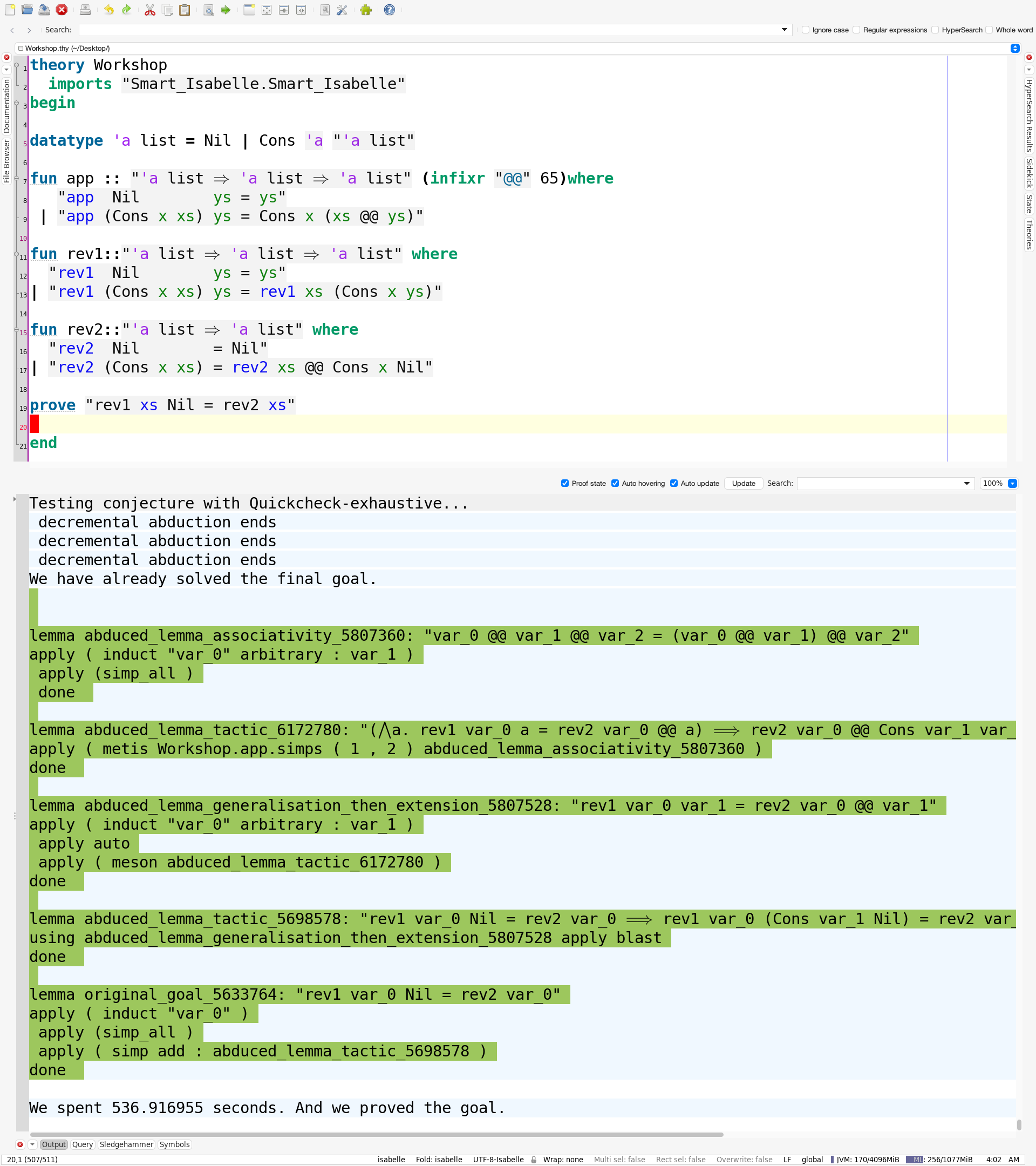}
    \caption{Proof script produced by the AbductionProver for the running example.}
    \label{fig:screenshot}
\end{figure}

To automatically address these three questions, this paper presents the AbductionProver.
Given a challenging proof goal, the AbductionProver keeps generating and proving useful auxiliary lemmas to prove the challenging goal.
For example, in our running case, it produces the proof script shown in Fig.~\ref{fig:screenshot}.

Our implementation is specific to Isabelle/HOL;
however, the underlying concepts presented in this paper are transferrable to other tactical theorem provers with similar ecosystems.

In the rest of the paper, we introduce the AbductionProver in a step-by-step manner.
That is, we start by presenting the default manual proof search model of Isabelle against a simpler problem in Section 1.
And gradually add components of the AbductionProver in each section to overcome each challenge. 
    The key to this reasoning is the two flags attached to nodes, completion status flag and contributivity status flag, and the distinction between two kinds of graphs: the abduction graph and the solution graph.

Contrary to other applications of artificial intelligence in theorem proving, the AbductionProver currently does not directly rely on machine learning or meta-heuristic approaches such as evolutionary computation.
Instead, it embodies a proof search framework, which could benefit from such approaches.

Although this paper primarily focuses on the design and architecture of the AbductionProver,
a working implementation has been developed on top of the PSL framework for Isabelle/HOL.
To facilitate reproducibility, future experimentation, and further extensions, the following public artifacts related to this work are available online:

\begin{itemize}[noitemsep]
    \item Demonstration video:
    \url{https://www.youtube.com/watch?v=rXU-lJxP_GI}

    \item PSL framework for Isabelle/HOL:
    \url{https://github.com/data61/PSL}
\end{itemize}

\section{From Implicit OR-tree Exploration To AND-OR graph Expansion}\label{sec:background}

\subsection{Tactic-Based Proof Search as Implicit OR-tree Exploration}\label{sec:ortree-tactic}

We start by reviewing the process of tactical theorem proving as an implicit OR-tree exploration using our running example from Section \ref{sec:intro}.

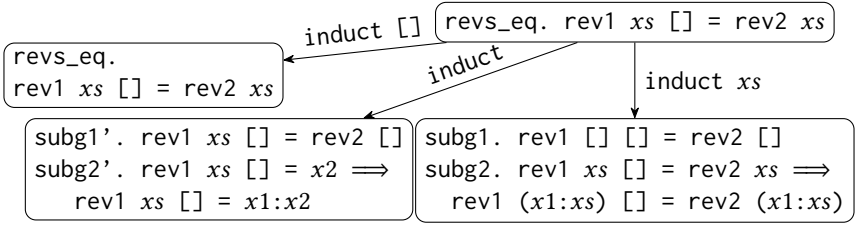
\begin{figure}[t]
\centering
\begin{tikzpicture}
[>={Stealth[round, scale=1.0]}, 
node distance=2cm, 
every edge/.style={draw, ->}, 
every edge quotes/.style = {auto, draw=none},
every node/.style={draw, ->, rounded corners, align=left}]
  \node[] (A) {\texttt{revs\_eq. rev1 $xs$ [] = rev2 $xs$}};
  \node[below=1.0cm of A, xshift=-5.5cm] (B) {    \texttt{subg1'. rev1 $xs$ [] = rev2 []}\\
     \texttt{subg2'. rev1 $xs$ [] = $x2$ $\Longrightarrow$}\\
     \texttt{\hspace{1.5em}rev1 $xs$ [] = $x1$:$x2$}
    };
  \node[below=1.0cm of A, xshift=0.0cm] (C)
    { \texttt{subg1. rev1 [] [] = rev2 []}\\
      \texttt{subg2. rev1 $xs$ [] = rev2 $xs$ $\Longrightarrow$}\\
      \hspace{1.0em}\texttt{rev1 ($x1$:$xs$) [] = rev2 ($x1$:$xs$)}
      };
  \node[below=0.0cm of A, xshift=-6.5cm] (D){\texttt{revs\_eq.}\\\texttt{rev1 $xs$ [] = rev2 $xs$}};
  \path
    (A) edge[sloped, "\texttt{induct}"] (B)
    (A) edge["\texttt{induct $xs$}"] (C)
    (A) edge[sloped, "\texttt{induct []}"] (D)
    ;
\end{tikzpicture}
\Description{The default computational model of Isabelle: Tactical theorem proving as an implicit OR-tree traversal.}
\caption{The default computational model of Isabelle: Tactical theorem proving as an implicit OR-tree traversal.}
\label{fig:ortree-tactic}
\end{figure}

Fig. \ref{fig:ortree-tactic} shows three scenarios of applying induction. 
The left scenario passes the empty list as the argument to the \texttt{induct} tactic,
the one in the middle applies structural induction on the $xs$ on right-hand side, and the right one applies structural induction on both occurrences of $xs$.
Fig. \ref{fig:ortree-tactic} also exhibits a few notable issues that are important to understand the design of the AbductionProver later in this paper.

First, this figure represents the tactical theorem proving as an implicit OR-tree traversal, which is the default computational model of Isabelle.
In this mode, multiple subgoals emerging from the previous step are grouped into one node.

Second, the left scenario did not make meaningful progress: the resulting subgoal is the same as the original goal.
Therefore, the choice of tactic argument is inappropriate.
Nevertheless, this subgoal is still logically provable,
posing a challenge in avoiding exponential blow-up of the search space.
We will discuss how the AbductionProver handles such cases in Section \ref{sec:beyond-abduction}.

Third, we can consider the tactic applications in the middle and on the right as instances of Modus Ponens (MP) with multiple premises.
For example, the application of the \texttt{induct} tactic ensures that if we prove the base-case and step-case, that is tantamount to proving the original goal:

\begin{prooftree}
  \AxiomC{}
  \RightLabel{\texttt{(induct xs)}}
  \UnaryInfC{\texttt{subg1} $\land{}$ \texttt{subg2} $\Longrightarrow{}$ \texttt{revs\_eq}}

  \AxiomC{\texttt{subg1} $\land{}$ \texttt{subg2}}

  \RightLabel{MP}
  \BinaryInfC{\texttt{revs\_eq}}
\end{prooftree}

\noindent
Generally, tactic applications can be seen as following using Modus Ponens:

\begin{prooftree}
  \AxiomC{}
  \RightLabel{\texttt{tactic}}
  \UnaryInfC{\texttt{sub-goals} $\Longrightarrow{}$ \texttt{goal}}

  \AxiomC{\texttt{sub-goals}}

  \RightLabel{MP-tac}
  \BinaryInfC{\texttt{goal}}
\end{prooftree}

The directions of the arrows differ between the figure and the derivation trees, and this requires explanation.
In the derivation trees, $\Longrightarrow$ represents logical implication: $\texttt{sub-goals} \Longrightarrow \texttt{goal}$ means that we must prove the $\texttt{goal}$ assuming the $\texttt{sub-goals}$.
In contrast, the arrows connecting nodes in Fig.~\ref{fig:ortree-tactic} indicate the direction of tree traversal, which is the standard notation for tree search.
Therefore, \singlebox{goal} $\longrightarrow{}$ \singlebox{sub-goal} denotes that we proceed to \singlebox{sub-goal} to prove \singlebox{goal}.

For problems at this difficulty level, we can often identify provably correct arguments for the \texttt{induct} tactic by executing a backtracking search over tactic combinations and argument candidates.
The backtracking search aligns well with the OR-tree traversal model shown in Fig. 
\ref{fig:ortree-tactic}, as it obtains final proof scripts by concatenating the labels on the edges from the root node to the leaf node representing proof completion \cite{psl}.
This approach effectively addressed \textsf{Q1} in Section \ref{sec:intro}, and we use it in our tool as explained in Section \ref{sec:try-to-prove}.
However, its efficacy was limited to simple problems that did not require explicit conjecturing.

\subsection{Tactic-Based Search as the Construction of an AND-OR Tree}\label{sec:andOR-tree-tactic}

\begin{figure}[t]
\centering
\begin{tikzpicture}
[>={Stealth[round, scale=1.5]}, 
node distance=2cm, 
every edge/.style={draw, ->}, 
every edge quotes/.style = {draw=none},
every node/.style={draw, ->, rounded corners, align=left}]
  \node[rounded corners] (A) {\verb|revs_eq|};

  \node[double, below=0.0cm of A, xshift=-4.0cm] (B){\verb|revs_eq|};
  \node[below=0.3cm of B] (B1){\verb|revs_eq|};
  \node[double, below=1.0cm of A, xshift=-1.5cm] (C) {\texttt{subg1' $\land$ subg2'}};
  \node[below=0.3cm of C, xshift=-1.0cm] (C1) {\texttt{subg1'}};
  \node[below=0.3cm of C, xshift=1.0cm] (C2) {\texttt{subg2'}};
  \node[double, below=1.0cm of A, xshift=2.5cm] (D) {\texttt{subg1 $\land$ subg2}};
  \node[below=0.3cm of D, xshift=-1.5cm] (D1)
    {\texttt{subg1}};
  \node[below=0.3cm of D, xshift=1.5cm] (D2)
    {\texttt{subg2}};
  \node[draw=none, below=0.0cm of A, xshift=4.0cm] (E) {...};
  \path
    (A) edge[auto, sloped, "\texttt{induct []}"] (B)
    (B) edge [] (B1)
    (A) edge[] node[pos=0.5, auto, sloped, draw=none] {\texttt{...}} (C)
    (C) edge [] (C1)
    (C) edge [] (C2)
    (A) edge[] node[draw=none, pos=0.1] {\hspace{5.5em}\texttt{induct $xs$}} (D)
    (A) edge[] node[draw=none, pos=0.4] {\hspace{8.5em}\texttt{simp only: subg1}} (D)
    (A) edge[] node[draw=none, pos=0.7] {\hspace{8.5em}\texttt{simp only: subg2}} (D)
    (A) edge[dashed] (D)
    (D) edge[](D1)
    (D) edge[](D2)
    (A) edge[dashed](E)
    ;
\end{tikzpicture}
\Description{Tactic application as an AND-OR tree construction.}
\caption{Tactic application as an AND-OR tree construction.}
\label{fig:AND-OR-tree-tactic}
\end{figure}
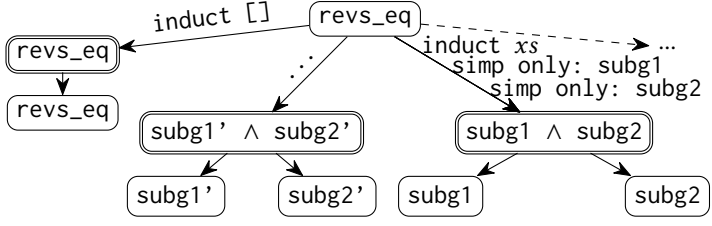

Another way to understand such computation is by considering it as an expansion of an AND-OR tree, as illustrated in Fig. \ref{fig:AND-OR-tree-tactic}. 
When depicting AND-OR structures in this paper, we use double-line borders for AND-nodes within AND-OR models to indicate that all their sub-structures must be proved.
In contrast, single-line borders represent OR-nodes to signify that only one of their sub-structures needs to be proven.

Such AND-OR trees might seem more intuitive to many ITP users, as they view each subgoal as an individual problem. 
However, the AND-OR tree model makes it challenging to keep track of which sub-trees need to be proved, especially after applying multiple tactics in sequence, causing the tree to grow deeper.
Also, this model deviates from the internal representation of proof goals in Isabelle and Lean, which are closer to the OR-tree representation, making it less suitable as the platform to execute the aforementioned backtracking search over tactic applications.

The AbductionProver combines the strengths of both representations, 
by performing a small-scale backtracking search over the OR-tree model while orchestrating large-scale proof automation using a data-structure inspired by the AND-OR tree model, as explained in Section \ref{sec:recursive-abduction}.

\subsection{One-Step Conjecturing}\label{sec:one-step}

Section \ref{sec:intro} demonstrated that, we often need to first establish intermediate lemmas as stepping stones to prove challenging theorems. 
Such intermediate lemmas have to satisfy the following conditions:

\begin{enumerate}[label=\textsf{C\arabic*:}]
    \item The lemmas must be useful for proving the challenging theorems.
    \item The lemmas must themselves be provable.
    \item The lemmas must be easier to prove than the challenging theorems.
\end{enumerate}
\noindent
For example, we can prove our running example with the following lemma.

\begin{alltt}
helper: "rev1 \(xs\) \(ys\) = rev2 \(xs\) + \(ys\)" sorry
subg2: "rev1 \(xs\) [] = rev2 xs \(\Longrightarrow\) rev1 (\(x1\):\(xs\)) [] = rev2 (\(x1\):\(xs\))"
apply(auto intro: helper) done
\end{alltt}
\noindent
where we naively assumed the correctness of the \verb|helper| lemma using the \verb|sorry| keyword.

Note that this paper's main focus is on our approach to identify useful lemmas out of many candidate conjectures, and we defer the description of conjecture generation to Section \ref{sec:explicit-conjecturing}, in contrast to other projects that focus on producing such auxiliary lemmas using machine learning techniques.

Similarly to the application of the \verb|induct| tactic in Section \ref{sec:ortree-tactic}, 
we consider this proof as an instance of Modus Ponens as follows:

\noindent
\begin{prooftree}
  \AxiomC{}
  \RightLabel{\texttt{(auto intro: helper)}}
  \UnaryInfC{\texttt{helper} $\Longrightarrow$ \texttt{subg2}}

  \AxiomC{}
  \RightLabel{\texttt{sorry}}
  \UnaryInfC{\texttt{helper}}

  \RightLabel{MP}
  \BinaryInfC{\texttt{subg2}}
\end{prooftree}
Based on this interpretation, this paper uses abductive reasoning to identify useful lemmas for many challenging theorems.
The key is to produce many candidate conjectures and filter less promising candidates using proof automation tools and counterexample finders.
Consider the following three candidate conjectures, $\bot$, \texttt{conj}, and $\top$, as follows:

\noindent
\begin{minipage}{0.25\textwidth}
\begin{prooftree}
  \AxiomC{$\bot \Longrightarrow$ \texttt{subg2}}
  \AxiomC{$\bot$}
  \BinaryInfC{\texttt{subg2}}
\end{prooftree}
\end{minipage}
\hfill
\begin{minipage}{0.45\textwidth}
\begin{prooftree}
  \AxiomC{\texttt{conj} $\Longrightarrow$ \texttt{subg2}}
  \AxiomC{\texttt{conj}}
  \RightLabel{MP-conj}
  \BinaryInfC{\texttt{subg2}}
\end{prooftree}
\end{minipage}
\hfill
\begin{minipage}{0.25\textwidth}
\begin{prooftree}
  \AxiomC{$\top \Longrightarrow$ \texttt{subg2}}
  \AxiomC{$\top$}
  \BinaryInfC{\texttt{subg2}}
\end{prooftree}
\end{minipage}

\noindent
where $\bot$ stands for \texttt{False} and $\top$ stands for \texttt{True} in the underlying non-minimal logic that admits the law of explosion.

The left derivation tree represents an extreme case where the candidate conjecture is a simple \texttt{False} statement. In this derivation tree, the left branch ($\bot \Longrightarrow$ \texttt{subg2}) can be easily proven because the underlying logic supports the law of explosion. However, it is impossible to prove the right branch ($\bot$). Generally, assuming \texttt{False} conjectures allows us to prove challenging goals, but such conjectures themselves cannot be proven.

The derivation tree on the right illustrates the opposite extreme, where the candidate conjecture is a simple \texttt{True} statement. In this tree, the right branch trivially holds, but proving the left branch is no easier than proving the original goal. This is because having \texttt{True} as an assumption provides no additional advantage in proving \texttt{subg2}.

The conjectures we aim for must not be equivalent to \texttt{False} and should also be useful for proving the final goals. To identify such conjectures, Nagashima \etal{} first apply Isabelle's proof tactic \texttt{fastforce} to filter in candidate conjectures 
that are provably useful to attack the final goals. 
Subsequently, they apply counterexample finders (\texttt{Quickcheck} and \texttt{Nitpick}) to filter out conjectures that are equivalent to \texttt{False}. 

This process generates conjectures that are provably useful but not demonstrably equivalent to \texttt{False}. However, it does not guarantee that the final criterion for intermediate lemmas — \textsf{C3} ease of proof — is satisfied. Only when we successfully prove the lemma using smaller computational resources can we confirm that the lemmas meet all three criteria.

\begin{figure}[t]
\centering
\begin{tikzpicture}
[>={Stealth[round, scale=1.5]}, 
every edge/.style={draw, ->}, 
every edge quotes/.style = {draw=none},
every node/.style={draw, ->, rounded corners}]
  \node[] (A) {\texttt{subg2}};
  \node[draw=gray, dashed, below=0.5cm of A, xshift=-5.0cm, text=gray] (B) {$\bot$};
  \node[below=0.8cm of A, align=center] (C)
    { \texttt{helper} };
  \node[draw=gray, dashed, below=0.5cm of A, xshift=5.0cm, text=gray] (D) {$\top$};
  \path
    (A) edge[draw=gray, dashed] node[pos=0.5, above, draw=none, sloped, text=gray] 
     {counterexample for $\bot$} (B)
    (A) edge["\texttt{auto intro: helper}"] (C)
    (A) edge[draw=gray, dashed] node[pos=0.5, above, draw=none, sloped, text=gray] {difficult to prove } (D)
    (A) edge[draw=none, dashed] node[pos=0.5, below, draw=none, sloped, text=gray] 
     {$\top \Longrightarrow$ \texttt{subg2}} (D);
\end{tikzpicture}
\Description{Tactic application as an OR-tree construction.}
\caption{Tactic application as an OR-tree construction.}
\label{fig:ortree-conjecture}
\end{figure}
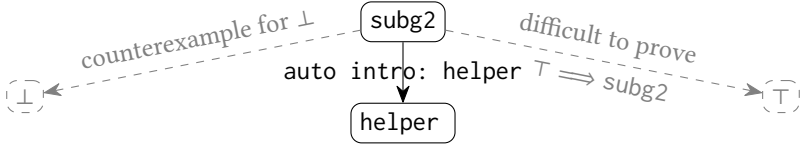

Similar to the case with tactics, this process can be viewed either as an OR-tree search, as illustrated in Fig. \ref{fig:ortree-conjecture}. In this figure, two dashed gray edges extend from \texttt{subg2} to dashed gray nodes: 
\dashedgraysinglebox{$\bot$} and \dashedgraysinglebox{$\top$},
indicating that the abductive reasoning process decided not to pursue these branches, with the reasons provided as the edge labels.

Based on this intuition, Nagashima \etal{} automated abductive reasoning as a backtracking search over an implicit OR-tree to identify useful conjectures.
They demonstrated the efficacy of this approach with a small example; however, its computational complexity increases rapidly when applied to challenging problems that require nested conjecturing.
Not only does the search space grow exponentially with the depth of the implicit OR-tree, but the same conjectures are also produced in different branches, causing duplication.
For instance, a common property, such as the associativity of the \verb|+| operator in this example, can be useful for proving many conjectures appearing in different branches.
This results in repeated production of this lemma and its proofs, which we address in Section \ref{sec:andor-graph}.

\subsection{Tactic Application as Implicit Conjecturing}\label{sec:tactic_as_conjecturing}

Attentive readers may have noticed the similarities between MP-tac and MP-conj, as well as between Fig. \ref{fig:ortree-tactic} and Fig. \ref{fig:ortree-conjecture}.
That is, both tactic applications and explicit conjecturing are instances of Modus Ponens. 

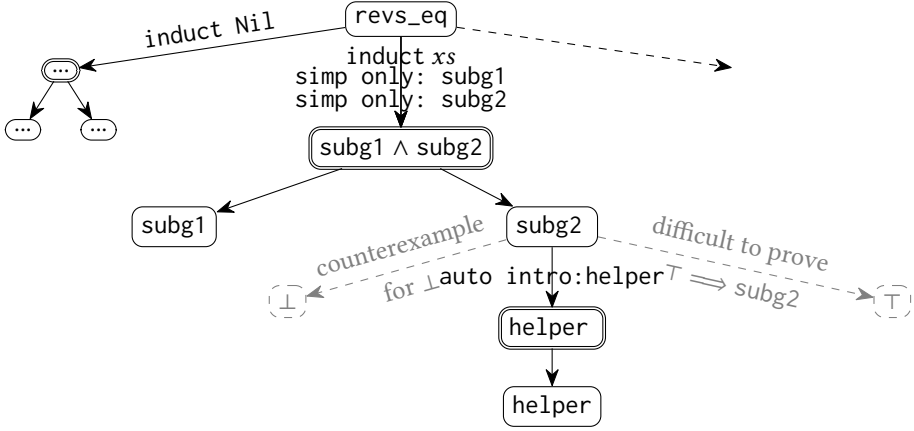
\begin{figure}[t]
\centering
\begin{tikzpicture}
[>={Stealth[round, scale=1.5]}, 
node distance=2cm, 
every edge/.style={draw, ->}, 
every edge quotes/.style = {draw=none},
every node/.style={draw, ->, rounded corners}]
  \node[] (A) {\verb|revs_eq|};
  \node[double, below=0.3cm of A, xshift=-4.5cm] (B) {...};
  \node[draw=none,below=0.3cm of A, xshift=4.5cm] (D) {};
  \node[double, below=1.2cm of A] (C)
    { \texttt{subg1} $\land$ \texttt{subg2}};
  \node[below=0.5cm of C, xshift=-3.0cm, align=center] (C1) {\texttt{subg1}};
  \node[below=0.5cm of C, xshift=2.0cm, align=center] (C2) { \texttt{subg2} };
  \node[below=0.5cm of B, xshift=-0.5cm] (B1) {...};
  \node[below=0.5cm of B, xshift=0.5cm] (B2) {...};
  \node[draw=gray, dashed, below=0.5cm of C2, xshift=-3.5cm, text=gray] (C21) {$\bot$};
  \node[double, below=0.8cm of C2, align=center] (C22) { \texttt{helper} };
  \node[draw=gray,dashed, below=0.5cm of C2, xshift=4.5cm, text=gray] (C23) {$\top$};
  \node[below=0.5cm of C22] (C221) {\texttt{helper}};
  \path
    (A) edge[auto, sloped, "\texttt{induct Nil}"] (B)
    (A) edge[] node[draw=none, pos=0.2] {\texttt{induct} $xs$} (C)
    (A) edge[] node[draw=none, pos=0.45] {\texttt{simp only: subg1}} (C)
    (A) edge[] node[draw=none, pos=0.7] {\texttt{simp only: subg2}} (C)
    (A) edge[dashed] (D)
    (B) edge[] (B1)
    (B) edge[] (B2)
    (C) edge[] (C1)
    (C) edge[] (C2)
    (C2) edge[draw=gray, dashed] node[pos=0.5, above, draw=none, sloped, text=gray] {counterexample} (C21)
    (C2) edge[draw=none, dashed] node[pos=0.5, below, draw=none, sloped, text=gray] {for $\bot$} (C21)
    (C2) edge["\texttt{auto intro:helper}"] (C22)
    (C2) edge[draw=gray, dashed] node[pos=0.5, above, draw=none, sloped, text=gray] {difficult to prove } (C23)
    (C2) edge[draw=none, dashed] node[pos=0.5, below, draw=none, sloped, text=gray] {$\top \Longrightarrow$ \texttt{subg2}} (C23)
    (C22) edge[] (C221)
    ;
\end{tikzpicture}
\Description{Tactic application and explicit conjecturing as an AND-OR tree construction.}
\caption{Tactic application and explicit conjecturing as an AND-OR tree construction.}
\label{fig:AND-OR-tree}
\end{figure}

Since they are based on the same rule, the AbductionProver sees \textit{tactic application as implicit conjecturing} and treats them in essentially the same way with adjustments detailed in Section \ref{sec:exp_tac}.
That is, instead of having separate trees for each proof goal appearing during large proof developments, we have one AND-OR tree that encompasses both tactic applications and conjecturing in a monolithic manner as shown in Fig. \ref{fig:AND-OR-tree}.

This way, the AbductionProver simplifies the question of when to introduce new lemmas, presented as \textsf{Q2} in Section \ref{sec:background}, into the straightforward task of deciding which node to expand next within a single structure, the Abduction Graph.
Although this may resemble the standard best-first search problem found in many other domains, there is a key difference: the goal of expanding nodes is to grow the AND-OR structure until it encompasses a sub-structure that represents a complete proof of the root node.

\subsection{From AND-OR Tree to AND-OR Graph}\label{sec:andor-graph}

As discussed in Section \ref{sec:andOR-tree-tactic}, 
the AND-OR tree model makes it challenging to keep track of which subtrees need to be proved compared to the OR-tree model, which represents a proof script as a single path from the root node to a leaf node, signifying the completion of the proof. Furthermore, the OR-tree model frames proof search as a procedure to identify a single leaf node, making it more aligned not only with Isabelle's default computation model but also with common AI problems in other fields. 

Despite these considerations, the AbductionProver is based on an AND-OR model because different intermediate goals often require the same conjectures as their stepping stones. For instance, the associativity of the \verb|+| operator is likely to be useful for solving many intermediate goals encountered while trying to prove \verb|revs_eq|.

To avoid duplicated lemmas,
we further switched from the AND-OR tree model to an AND-OR graph model,
since the graph model better represents situations where multiple sub-goals can be proved using the same intermediate lemmas.

\begin{figure}[t]
\centering
\begin{tikzpicture}
[>={Stealth[round, scale=1.5]}, 
node distance=2cm, 
every edge/.style={draw, ->}, 
every edge quotes/.style = {draw=none},
every node/.style={draw, ->, rounded corners, align=center}]
  \node[] (A) {\verb|revs_eq|};
  \node[double, below=1.2cm of A] (C)
    { \texttt{subg1} $\land$ \texttt{subg2}};
  \node[below=0.5cm of C, xshift=-1.0cm] (C1) {\texttt{subg1}};
  \node[below=0.5cm of C, xshift=1.0cm] (C2) { \texttt{subg2} };
  \node[double, right=3.0cm of C2] (C22) { \texttt{helper} };
  \node[below=0.0cm of C22, xshift=2.0cm] (C221) {\texttt{helper}};
  \path
    (A) edge[] node[draw=none, pos=0.2] {\texttt{induct} $xs$\texttt{}} (C)
    (A) edge[] node[draw=none, pos=0.45] {\texttt{simp only: subg1}} (C)
    (A) edge[] node[draw=none, pos=0.7] {\texttt{simp only: subg2}} (C)
    (C) edge[] (C1)
    (C) edge[] (C2)
    (C2) edge[] node[pos=0.5, below, draw=none, sloped] {\texttt{auto intro:helper}} (C22)
    (C22) edge[] (C221)
    (A) edge[bend left=10, auto] node[draw=none, sloped, above] {\texttt{metis helper rev1.simps}} (C22)
    (A) edge[bend left=10, auto] node[draw=none, sloped, below] {\texttt{rev2.elims}} (C22)
    ;
\end{tikzpicture}
\Description{Tactic application and conjecturing as an AND-OR graph construction.}
\caption{Tactic application and conjecturing as an AND-OR graph construction.}
\label{fig:andor-graph}
\end{figure}
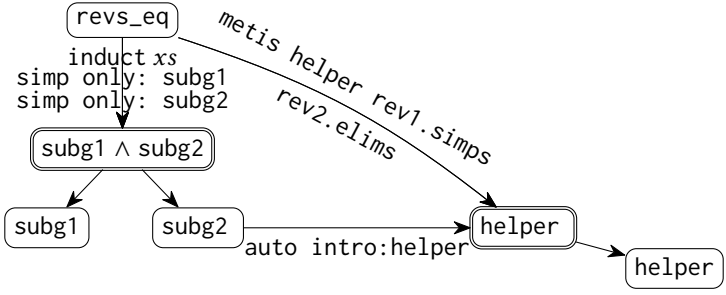

Fig. \ref{fig:andor-graph} illustrates multiple dependencies on a single lemma using the running example. 
Specifically, this figure shows that \singlebox{revs\_eq}
can be directly proved by applying the \texttt{metis} tactic
once \doublebox{helper} is established.
The \texttt{metis} tactic takes three arguments:
\texttt{helper}, \texttt{rev1.simps}, and \texttt{rev2.elims}.
The first argument (\texttt{helper}) appears in the graph as \doublebox{helper} and \singlebox{helper}.

On the contrary, \texttt{rev1.simps} and \texttt{rev2.elims} are absent from the graph; they are registered in the background proof context
when the relevant functions (\texttt{rev1} and \texttt{rev2}) are defined.
These lemmas are made available as contextual information for tools like \texttt{sledgehammer}.
Overall, the AbductionProver uses the proof context in a restrictive manner:
it tries to register as few additional lemmas as possible in the background context
so that it can explicitly handle the inter-dependencies among auxiliary lemmas.

Having introduced the benefits of using the AND-OR graph model for proof search involving multi-step conjecturing, we now invite readers to reflect on the following question: \textit{Must such AND-OR graphs be acyclic, or can they be cyclic}?

\section{Reasoning Over Abduction Graph}\label{sec:abduction-graph}

Before answering this question, we examine how to reason over abduction graphs.
The two major questions to address when using abduction graphs are as follows:
\begin{enumerate}
\item \textsf{Completion:} Which nodes have already been proved?
\item \textsf{Contributivity:} Which nodes should still be attempted to prove?
\end{enumerate}
As noted in Section \ref{sec:andor-graph}, tracking which nodes have already been proved is challenging with AND-OR trees and even more so with AND-OR graphs.
Therefore, we first determine how to verify whether
\singlebox{revs\_eq}
is completely proved in our running example.

\subsection{Completed Nodes: Which Nodes Are Already Proved?}\label{sec:completion}

At the stage shown in Fig. \ref{fig:andor-graph}, none of the OR-nodes have been proved yet, including the root node (\singlebox{revs\_eq}). 
For the root node, the graph only indicates that proving either
\doublebox{subg1 $\land$ subg2}
or
\doublebox{helper}
would be equivalent to proving
\singlebox{revs\_eq}. 
However, neither of these and-nodes have been proved.

Fig. \ref{fig:subg1-proved} and Fig. \ref{fig:helper-proved} illustrate how this situation changes when
\singlebox{subg1}
and
\singlebox{helper}
are proved, respectively.
The specific steps taken to prove these or-nodes are not the focus; rather, we examine how proving them affects the proof completion status of their ancestral nodes at this stage.

\begin{figure}[t]
\begin{minipage}{0.48\textwidth}
\centering
\begin{tikzpicture}
[>={Stealth[round, scale=1.5]}, 
node distance=2cm, 
every edge/.style={draw, ->}, 
every edge quotes/.style = {draw=none},
every node/.style={draw, ->, rounded corners, align=center}]
  \node[] (A) {\verb|revs_eq|};
  \node[double, below=0.5cm of A] (C)
    { \texttt{subg1} $\land$ \texttt{subg2}};
  \node[fill=blue!25, below=0.5cm of C, xshift=-1.0cm] (C1) {\texttt{subg1}};
  \node[draw=none, below=-0.1cm of C1] (C11) {\checkmark};
  \node[below=0.5cm of C, xshift=1.0cm] (C2) { \texttt{subg2} };
  \node[double, below=0.5cm of C2, xshift=1.0cm] (C22) { \texttt{helper} };
  \node[below=0.0cm of C22, xshift=-2.0cm] (C221) {\texttt{helper}};
  \path
    (A) edge[] (C)
    (C) edge[] (C1)
    (C1) edge[bend left=25, draw=blue, dashed] node[pos=0.5, left, draw=none, text=blue] {\ding{172}} (C)
    (C) edge[] (C2)
    (C2) edge[] (C22)
    (C22) edge[] (C221)
    (A) edge[bend left=35] (C22)
    ;
\end{tikzpicture}
\caption{Upward completion check due to \texttt{subg1}.}
\label{fig:subg1-proved}
\end{minipage}
\hfill
\begin{minipage}{0.48\textwidth}
\centering
\begin{tikzpicture}
[>={Stealth[round, scale=1.5]}, 
node distance=2cm, 
every edge/.style={draw, ->}, 
every edge quotes/.style = {draw=none},
every node/.style={draw, ->, rounded corners, align=center}]
  \node[] (A) {\verb|revs_eq|};
  \node[double, below=0.5cm of A] (C)
    { \texttt{subg1} $\land$ \texttt{subg2}};
  \node[below=0.5cm of C, xshift=-1.0cm] (C1) {\texttt{subg1}};
  \node[below=0.5cm of C, xshift=1.0cm] (C2) { \texttt{subg2} };
  \node[double, below=0.5cm of C2, xshift=1.0cm] (C22) { \texttt{helper} };
  \node[fill=blue!25, below=0.0cm of C22, xshift=-2.0cm] (C221) {\texttt{helper}};
  \node[draw=none, below=-0.1cm of C221] (C2211) {\checkmark};
  \path
    (A) edge[] (C)
    (C) edge[] (C1)
    (C) edge[] (C2)
    (C2) edge[bend left=25, draw=blue, dashed] node[pos=0.5, below, draw=none, text=blue] {\ding{174}} (C)
    (C2) edge[] (C22)
    (C22) edge[bend left=25, draw=blue] node[pos=0.5, left, draw=none, text=blue] {\ding{173}} (C2)
    (C22) edge[] (C221)
    (A) edge[bend left=35] (C22)
    (C22) edge[bend right=60, draw=blue] node[pos=0.5, right, draw=none, text=blue] {\ding{173}}(A)
    (C221) edge[bend left=25, draw=blue] node[pos=0.5, left, draw=none, text=blue] {\ding{172}}(C22)
    ;
\end{tikzpicture}
\Description{\texttt{helper} newly proved.}
\caption{Upward completion check due to \texttt{helper}.}
\label{fig:helper-proved}
\end{minipage}

\end{figure}

In Fig.\ref{fig:subg1-proved}, even though we prove
\singlebox{subg1},
this does not immediately affect other nodes, since the only parent AND-node
(\doublebox{subg1 $\land$ subg2})
still has an unproved sub-goal (\singlebox{subg2}).
The dashed arrow from \singlebox{subg1} to \doublebox{subg1 $\land$ subg2}
indicates that the proof of the \singlebox{subg1} fails to update the proof status of
\doublebox{subg1 $\land$ subg2}.
As indicated by this arrow, the process works primarily in a bottom-up manner.
However, when evaluating the proof completion status of an ancestral AND-node
such as \doublebox{subg1 $\land$ subg2},
it must look downward exactly one layer to gather the proof statuses of all its immediate child OR-nodes
(\singlebox{subg1} and
\singlebox{subg2}).
To facilitate this, the graph must maintain the current proof completion status of each OR-node.

Fig. \ref{fig:helper-proved}, on the other hand, 
shows how completion check propagates
when we prove 
\singlebox{helper} without proving
\singlebox{subg1}.
In this scenario, proving \singlebox{helper} automatically proves its parental AND-node (\doublebox{helper}), 
because \singlebox{helper} is the only child of \doublebox{helper}.
And this in turn is tantamount to prove \singlebox{revs\_eq} and \singlebox{subg2}.
However, \singlebox{subg2} does not change the status of \doublebox{subg1 $\land$ subg2}, since \singlebox{subg1} is not proved.

These figures demonstrated how to determine whether branch nodes are considered proved,
assuming leaf OR-nodes are proved.
While we defer the discussion of proving leaf nodes to Section \ref{sec:try-to-prove},
we now introduce the notions of \textit{completed} nodes and \textit{atomic} proofs to distinguish the efforts required to prove branch nodes and leaf nodes.

In general,
a leaf node is said to have an \textit{atomic} proof script if its proof does not rely on other nodes in the graph,
whereas a branch node is considered \textit{completed}
if the graph contains enough labeled edges and atomically proved leaf nodes
to establish its proof.
On the contrary, if a node is not completed, the node is said to be \textit{uncompleted}.

Specifically, an OR-node in an abduction graph is \textit{completed} if it satisfies one of the following conditions:
\begin{enumerate}[label=\textsf{D\arabic*:}]
\item The OR-node has an \textit{atomic} proof.
\item The OR-node points to at least one AND-node that is \textit{completed}.
\end{enumerate}
\noindent
In contrast, an AND-node in an abduction graph is \textit{completed} if it satisfies the following condition:
\begin{enumerate}[label=\textsf{E\arabic*:}]
\item The AND-node points to OR-nodes, all of which are \textit{completed}.
\end{enumerate}

In Fig. \ref{fig:helper-proved}, for example, 
\doublebox{subg1 $\land$ subg2} remains uncompleted,
since it has an uncompleted child OR-node (\singlebox{subg1}).
However, should we still complete
\singlebox{subg1} for
\doublebox{subg1 $\land$ subg2}
after proving
\singlebox{helper}?

\subsection{Contributive Nodes: Which Nodes Are Still Worth Proving?}\label{sec:contributive}

To determine whether a node should still be completed,
we need to know if it can still contribute to our objective:
completing the root node.

From this perspective, Fig. \ref{fig:after_proving_helper} and Fig. \ref{fig:not_worth_proving-after-helper} check if
we should complete
\singlebox{subg1} and
\doublebox{subg1 $\land$ subg2}
after proving
\singlebox{helper}.
First, Fig. \ref{fig:after_proving_helper} highlights in blue the nodes that are considered completed after proving
\singlebox{helper} in Fig. \ref{fig:helper-proved}.
Then, Fig. \ref{fig:not_worth_proving-after-helper} marks the nodes that remain uncompleted but are no longer necessary for achieving our goal with dashed boundaries.

That is, following the direction of arrows from the root node:
Since the root node is already completed, there is no need to complete its child node,
\doublebox{subg1 $\land$ subg2},
even though it has not been proved yet,
as illustrated by the edge labeled with \textcolor{blue}{\ding{172}}.

This process propagates downwards:
Since
\doublebox{subg1 $\land$ subg2}
is no longer necessary, there is no need to prove its child node,
\singlebox{subg1} either, as shown by the edge labeled with \textcolor{blue}{\textcircled{\textcolor{blue}{2}}}.

\begin{figure}[t]
\begin{minipage}{0.48\textwidth}
\centering
\begin{tikzpicture}
[>={Stealth[round, scale=1.5]}, 
node distance=2cm, 
every edge/.style={draw, ->}, 
every edge quotes/.style = {draw=none},
every node/.style={draw, ->, rounded corners, align=center}]
  \node[fill=blue!25] (A) {\verb|revs_eq|};
  \node[double, below=0.5cm of A] (C)
    { \texttt{subg1} $\land$ \texttt{subg2}};
  \node[below=0.5cm of C, xshift=-1.0cm] (C1) {\texttt{subg1}};
  \node[fill=blue!25, below=0.5cm of C, xshift=1.0cm] (C2) { \texttt{subg2} };
  \node[fill=blue!25, double, below=0.5cm of C2, xshift=1.0cm] (C21) { \texttt{helper} };
  \node[fill=blue!25, below=0.0cm of C21, xshift=-2.0cm] (C211) {\texttt{helper}};
  \node[draw=none, below=-0.1cm of C211] (C2111) {\checkmark};
  \path
    (A) edge[] (C)
    (C) edge[] (C1)
    (C) edge[] (C2)
    (C2) edge[] (C21)
    (C21) edge[] (C211)
    (A) edge[bend left=35] (C21)
    ;
\end{tikzpicture}
\caption{After completion check due to \texttt{helper}.}
\label{fig:after_proving_helper}
\end{minipage}
\hfill
\begin{minipage}{0.48\textwidth}
\centering
\begin{tikzpicture}
[>={Stealth[round, scale=1.5]}, 
node distance=2cm, 
every edge/.style={draw, ->}, 
every edge quotes/.style = {draw=none},
every node/.style={draw, ->, rounded corners, align=center}]
  \node[fill=blue!25] (A) {\verb|revs_eq|};
  \node[double, below=0.5cm of A, dashed] (C)
    { \texttt{subg1} $\land$ \texttt{subg2}};
  \node[below=0.5cm of C, xshift=-1.0cm, dashed] (C1) {\texttt{subg1}};
  \node[fill=blue!25, below=0.5cm of C, xshift=1.0cm] (C2) { \texttt{subg2} };
  \node[fill=blue!25, double, below=0.5cm of C2, xshift=1.0cm] (C21) { \texttt{helper} };
  \node[fill=blue!25, below=0.0cm of C21, xshift=-2.0cm] (C211) {\texttt{helper}};
  \node[draw=none, below=-0.1cm of C211] (C2111) {\checkmark};
  \path
    (A) edge[] (C)
    (A) edge[draw=blue, bend right=45] node[pos=0.5, left, draw=none, text=blue] {\ding{172}}(C)
    (C) edge[] (C1)
    (C) edge[draw=blue, bend right=45] node[pos=0.5, left, draw=none, text=blue] {\ding{173}}(C1)
    (C) edge[] (C2)
    (C2) edge[] (C21)
    (C21) edge[] (C211)
    (A) edge[bend left=35] (C21)
    ;
\end{tikzpicture}
\Description{\texttt{subg1} $\land$ \texttt{subg2} not needed.}
\caption{Downward contributivity check from the root.\\
\texttt{subg1} $\land$ \texttt{subg2} and \texttt{subg1} are no longer contributive.}
\label{fig:not_worth_proving-after-helper}
\end{minipage}
\end{figure}

In this particular case, it was straightforward to determine the unworthiness of these two nodes, since both nodes 
(\doublebox{subg1 $\land$ subg2} and
\singlebox{subg1})
are pointed by only one parental node.
However, if multiple OR-nodes point to one AND-node, completing one of the parent OR-nodes does not always eliminate the need to complete the AND-node.

Fig. \ref{fig:after-proving-repleh} and Fig. \ref{fig:not_worth_proving-after-repleh} illustrate this situation using an alternative auxiliary lemma, \texttt{repleh}, to prove \texttt{subg2}.
First, Fig. \ref{fig:after-proving-repleh} shows that proving the OR-node
\singlebox{repleh}
is sufficient to complete
\doublebox{repleh}
and
\singlebox{subg2}.
This is why these three nodes are highlighted in blue in Fig. \ref{fig:not_worth_proving-after-repleh}.
However, this does not eliminate the need for
\doublebox{helper}
in Fig. \ref{fig:not_worth_proving-after-repleh},
because
\doublebox{helper}
is pointed to by
\singlebox{revs\_eq},
and 
\singlebox{revs\_eq}
itself still needs to be completed.
Then, since the necessity for \doublebox{helper} has not changed this time, there is no need to check the necessity for its child node, 
\singlebox{helper}.

Generally, a node is considered \textit{non-contributive} 
if its completion does not contribute to the completion of the root node.
On the contrary, if its completion does contribute to the completion of the root node, the node is said to be \textit{contributive}.

Specifically, an OR-node becomes \textit{non-contributive} when it satisfies at least one of the following conditions:

\begin{enumerate}[label=\textsf{F\arabic*:}]
    \item The OR-node is already \textit{completed}.
    \item Both of the following two conditions hold:
\begin{itemize}
    \item The OR-node is not the root node.
    \item The OR-node is not pointed to by any \textit{contributive} AND-node.
\end{itemize}
\end{enumerate}
\noindent
In contrast, an AND-node becomes \textit{non-contributive}
when it satisfies at least one of the following conditions:
\begin{enumerate}[label=\textsf{G\arabic*:}]
    \item The AND-node is already \textit{completed}.
    \item The AND-node is not pointed to by any \textit{contributive} OR-node.
\end{enumerate}

\noindent
Following De Morgan's laws,
an OR-node becomes \textit{contributive} when it satisfies both of the following conditions:

\begin{enumerate}[label=\textsf{H\arabic*:}]
    \item The OR-node is still \textit{uncompleted}.
    \item At least one of the following two conditions holds:
\begin{itemize}
    \item The OR-node is the root node.
    \item The OR-node is pointed to by at least one \textit{contributive} AND-node.
\end{itemize}
\end{enumerate}
\noindent
In contrast, an AND-node becomes \textit{contributive}
when it satisfies both of the following condition:
\begin{enumerate}[label=\textsf{I\arabic*:}]
    \item The AND-node is still \textit{uncompleted}.
    \item The AND-node is pointed to by at least one \textit{contributive} OR-node.
\end{enumerate}

\begin{figure}[t]
\begin{minipage}{0.48\textwidth}
\centering
\begin{tikzpicture}
[>={Stealth[round, scale=1.5]}, 
node distance=2cm, 
every edge/.style={draw, ->}, 
every edge quotes/.style = {draw=none},
every node/.style={draw, ->, rounded corners, align=center}]
  \node[] (A) {\verb|revs_eq|};
  \node[double, below=0.5cm of A, align=center] (C)
    { \texttt{subg1} $\land$ \texttt{subg2}};
  \node[below=0.5cm of C, xshift=-1.0cm] (C1) {\texttt{subg1}};
  \node[below=0.5cm of C, xshift=1.0cm] (C2) { \texttt{subg2} };
  \node[double, below=0.5cm of C2, xshift=-1.0cm] (C21) { \texttt{repleh} };
  \node[below=0.5cm of C21, fill=blue!25] (C211) {\texttt{repleh}};
  \node[double, below=0.5cm of C2, xshift=1.0cm] (C22) { \texttt{helper} };
  \node[below=0.5cm of C22] (C221){\texttt{helper}};
  \node[draw=none, below=-0.1cm of C211] (C2111) {\checkmark};
  \path
    (A) edge[] (C)
    (C) edge[] (C1)
    (C) edge[] (C2)
    (C2) edge[draw=blue, bend left=40, dashed] node[pos=0.5, left, draw=none, text=blue] {\ding{174}} (C)
    (C2) edge[] (C21)
    (C21) edge[draw=blue, bend left=40] node[pos=0.5, left, draw=none, text=blue] {\ding{173}} (C2)
    (C2) edge[] (C22)
    (C21) edge[] (C211)
    (C211) edge[draw=blue, bend left=40] node[pos=0.5, left, draw=none, text=blue] {\ding{172}} (C21)
    (C22) edge[] (C221)
    (A) edge[bend left=35] (C22)
    ;
\end{tikzpicture}
\caption{Upward completion check due to \texttt{repleh}.}
\label{fig:after-proving-repleh}
\end{minipage}
\hfill
\begin{minipage}{0.48\textwidth}
\centering

\begin{tikzpicture}
[>={Stealth[round, scale=1.5]}, 
node distance=2cm, 
every edge/.style={draw, ->}, 
every edge quotes/.style = {draw=none},
every node/.style={draw, ->, rounded corners, align=center}]
  \node[] (A) {\verb|revs_eq|};
  \node[double, below=0.5cm of A] (C)
    { \texttt{subg1} $\land$ \texttt{subg2}};
  \node[below=0.5cm of C, xshift=-1.0cm] (C1) {\texttt{subg1}};
  \node[below=0.5cm of C, xshift=1.0cm, fill=blue!25] (C2) { \texttt{subg2} };
  \node[double, below=0.5cm of C2, xshift=-1.0cm, fill=blue!25] (C21) { \texttt{repleh} };
  \node[below=0.5cm of C21, fill=blue!25] (C211) {\texttt{repleh}};
  \node[double, below=0.5cm of C2, xshift=1.0cm] (C22) { \texttt{helper} };
  \node[below=0.5cm of C22] (C221) {\texttt{helper}};
  \node[draw=none, below=-0.1cm of C211] (C2111) {\checkmark};
  \path
    (A) edge[] (C)
    (C) edge[] (C1)
    (C) edge[] (C2)
    (C2) edge[] (C21)
    (C2) edge[] (C22)
    (C2) edge[draw=blue, bend right=25, dashed] node[pos=0.5, left, draw=none, text=blue] {\ding{172}} (C22)
    (C21) edge[] (C211)
    (C22) edge[] (C221)
    (A) edge[bend left=35] (C22)
    ;
\end{tikzpicture}
\Description{\texttt{helper} still needed.}
\caption{Downward contributivity check from the root.\\
\texttt{helper} is still contributive.}
\label{fig:not_worth_proving-after-repleh}
\end{minipage}
\end{figure}


\subsection{Upwards Completion Check Before Downwards Contributivity Check.}\label{sec:completion_contributivity}

The conditions presented in Sections \ref{sec:completion} and \ref{sec:contributive} imply that updating a node requires knowledge of the current status of its neighboring nodes.
For this reason, we must store the status of contributive nodes as required by the update process.

Regarding the dependencies among nodes in maintaining up-to-date status information, the criteria from these sections indicate the following:

\begin{enumerate} 
\item A node’s completion status depends on the completion status of its child nodes. 
\item A node’s completion status is independent of the contributivity status of any node. 
\item A node’s contributivity status depends on the contributivity status of its parent nodes. 
\item A node’s contributivity status depends on its own completion status. \end{enumerate}

\noindent
From these, we derive the following design principles for the AbductionProver:

\begin{enumerate} 
\item Completion status changes must be propagated upwards to its ancestors. 
\item Propagation of contributivity status changes does not affect completion statuses. 
\item Contributivity status changes must be propagated downwards to its descendants. 
\item Updates to completion statuses should be finalized before updating contributivity statuses. 
\end{enumerate}

Now that we know status updates propagate upward for completion status and downward for contributivity status,
does this imply that circular dependencies must be disallowed in abduction graphs to prevent infinite loops?

\subsection{Cyclic Abduction Graph and Acyclic Solution Graph}

\begin{figure}[t]
\begin{tikzpicture}
[>={Stealth[round, scale=1.5]}, 
node distance=2cm, 
every edge/.style={draw, ->}, 
every edge quotes/.style = {draw=none},
every node/.style={draw, ->, rounded corners, align=center}]
  \node[] (A) {\verb|revs_eq|};
  \node[double, below=0.5cm of A] (C) {\texttt{subg1} $\land$ \texttt{subg2}};
  \node[below=0.3cm of C, xshift=-1.0cm] (C1) {\texttt{subg1}};
  \node[below=0.3cm of C, xshift=3.0cm] (C2) {\texttt{subg2} };
  \node[double, below=0.3cm of C2, xshift=-4.5cm] (C21) {\texttt{repleh}};
  \node[below=0.5cm of C21] (C211) {\texttt{repleh}};
  \node[double, below=0.3cm of C2, xshift=4.5cm] (C22) {\texttt{helper}};
  \node[below=0.5cm of C22] (C221){\texttt{helper}};
  \path
    (A) edge[] (C)
    (C) edge[] (C1)
    (C) edge[] (C2)
    (C2) edge[] node[above, sloped, draw=none]{\texttt{auto intro:repleh}}(C21)
    (C2) edge[] node[above, sloped, draw=none]{\texttt{auto intro:helper}}(C22)
    (C21) edge[] (C211)
    (C22) edge[] (C221)
    (C211) edge[] node[pos=0.3, below, sloped, draw=none]{\texttt{auto simp:helper}} (C22)
    (C221) edge[] node[pos=0.675, above, sloped, draw=none]{\texttt{auto simp:repleh}} (C21)
    (A) edge[bend left=20] (C22)
    ;
\end{tikzpicture}
\Description{Cyclic abduction graph.}
\caption{Cyclic abduction graph.}
\label{fig:cycle}
\end{figure}
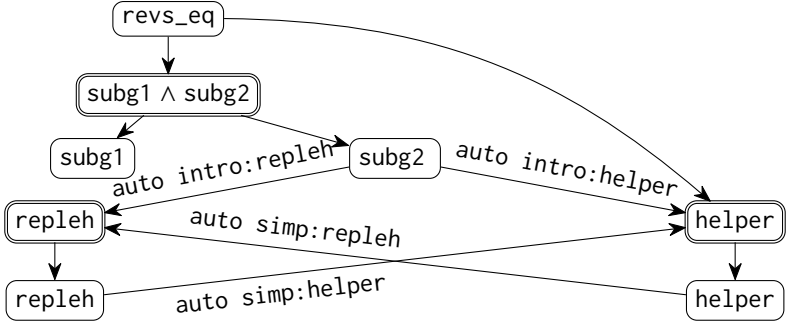

Fig.\ref{fig:cycle} is an example abduction graph for our running example, which shows a cyclic dependency between
\singlebox{helper}
and
\singlebox{repleh}.
Specifically, Fig. \ref{fig:cycle} contains two additional labeled edges
compared to Fig. \ref{fig:after-proving-repleh}.
Note the following two key differences from Fig. \ref{fig:after_proving_helper}.
First, unlike Fig. \ref{fig:after-proving-repleh}, Fig. \ref{fig:cycle} assumes that neither 
\singlebox{helper} nor
\singlebox{repleh}
has been proved yet.
Second, Fig. \ref{fig:cycle} has two additional edges, which makes the graph cyclic.

We added the edge from \singlebox{helper} to \doublebox{repleh}, because if we assume \doublebox{repleh} we can prove \singlebox{helper} as follows:
\begin{alltt}
repleh: "rev2 \(xs\) + \(ys\) = rev1 \(xs\) \(ys\)" sorry
helper: "rev1 \(xs\) \(ys\) = rev2 \(xs\) + \(ys\)" apply (auto simp:repleh) done
\end{alltt}
\noindent
whereas we added the edge from \singlebox{repleh} to \doublebox{helper}, because if we assume \doublebox{helper} we can prove \singlebox{repleh} as follows:
\begin{alltt}
helper: "rev1 \(xs\) \(ys\) = rev2 \(xs\) + \(ys\)" sorry
repleh: "rev2 \(xs\) + \(ys\) = rev1 \(xs\) \(ys\)" apply (auto simp:helper) done
\end{alltt}
Both proofs are logically valid under the respective assumptions axiomatically introduced by the \texttt{sorry} keyword.
Now, we examine how to reason over this cyclic graph by considering the completion update for two cases.

\paragraph{Status Updates After Proving \texttt{helper}.}

\begin{figure}[t]
\begin{minipage}{0.48\textwidth}
\centering
\begin{tikzpicture}
[>={Stealth[round, scale=1.5]}, 
node distance=2cm, 
every edge/.style={draw, ->, text=blue}, 
every edge quotes/.style = {draw=none},
every node/.style={draw, ->, rounded corners, align=center}]
  \node[] (A) {\verb|revs_eq|};
  \node[double, below=0.5cm of A] (C) {\texttt{subg1} $\land$ \texttt{subg2}};
  \node[below=0.5cm of C, xshift=-1.0cm] (C1) {\texttt{subg1}};
  \node[below=0.5cm of C, xshift=1.0cm] (C2) {\texttt{subg2} };
  \node[double, below=0.5cm of C2, xshift=-1.0cm] (C21) {\texttt{repleh} };
  \node[below=0.5cm of C21] (C211) {\texttt{repleh}};
  \node[double, below=0.5cm of C2, xshift=1.0cm] (C22) {\texttt{helper} };
  \node[below=0.5cm of C22, fill=blue!25] (C221){\texttt{helper}};
  \node[draw=none, below=-0.1cm of C221] (C2211) {\checkmark};
  \path
    (A) edge[] (C)
    (C) edge[] (C1)
    (C) edge[] (C2)
    (C2) edge[] (C21)
    (C2) edge[] (C22)
    (C21) edge[] (C211)
    (C22) edge[] (C221)
    (C22) edge[draw=blue, bend left=40] node[pos=0.3, left, draw=none] {\ding{173}} (C2)
    (C211) edge[] (C22)
    (C22) edge[draw=blue, bend left=20] node[pos=0.75, below, draw=none] {\ding{173}} (C211)
    (C221) edge[] (C21)
    (C221) edge[draw=blue, bend right=40] node[pos=0.5, right, draw=none] {\ding{172}} (C22)
    (C2) edge[draw=blue, bend left=40, dashed] node[pos=0.5, left, draw=none] {\ding{174}} (C)
    (C211) edge[draw=blue, bend left=40] node[pos=0.5, left, draw=none] {\ding{174}} (C21)
    (A) edge[bend left=35] (C22)
    (C22) edge[bend right=60, draw=blue] node[pos=0.5, right, draw=none] {\ding{173}} (A)
    ;
\end{tikzpicture}
\caption{\texttt{helper} newly proved.}
\label{fig:update_cycle_helper}
\end{minipage}
\hfill
\begin{minipage}{0.48\textwidth}
\centering
\begin{tikzpicture}
[>={Stealth[round, scale=1.5]}, 
node distance=2cm, 
every edge/.style={draw=gray, ->}, 
every edge quotes/.style = {draw=none},
every node/.style={draw, ->, rounded corners, align=center, draw=gray}]
  \node[fill=blue!25, draw=black] (A) {\verb|revs_eq|};
  \node[text=gray, dashed, double, below=0.5cm of A] (C) {\texttt{subg1} $\land$ \texttt{subg2}};
  \node[text=gray, dashed, below=0.5cm of C, xshift=-1.0cm] (C1) {\texttt{subg1}};
  \node[text=gray, dashed, below=0.5cm of C, xshift=1.0cm] (C2) { \texttt{subg2} };
  \node[text=gray, dashed, double, below=0.5cm of C2, xshift=-1.0cm] (C21) { \texttt{repleh} };
  \node[text=gray, dashed, below=0.5cm of C21] (C211) {\texttt{repleh}};
  \node[double, draw=black, below=0.5cm of C2, xshift=1.0cm, fill=blue!25] (C22) { \texttt{helper}};
  \node[draw=black, below=0.5cm of C22, fill=blue!25] (C221){\texttt{helper}};
  \node[draw=none, below=-0.1cm of C221] (C2211) {\checkmark};
  \path
    (A) edge[dashed] (C)
    (C) edge[dashed] (C1)
    (C) edge[dashed] (C2)
    (C2) edge[dashed] (C21)
    (C2) edge[dashed] (C22)
    (C21) edge[dashed] (C211)
    (C22) edge[draw=black] (C221)
    (C211) edge[dashed] (C22)
    (C221) edge[dashed] (C21)
    (A) edge[draw=black, bend left=35] (C22)
    ;
\end{tikzpicture}
\Description{Solution graph in Fig. \ref{fig:update_cycle_helper}.}
\caption{Solution graph in Fig. \ref{fig:update_cycle_helper}.}
\label{fig:solution_graph_helper}
\end{minipage}
\end{figure}

Fig. \ref{fig:update_cycle_helper} shows the case where we find an atomic proof for
\singlebox{helper}.
The atomic proof for \singlebox{helper} makes this leaf OR-node completed.
Since \singlebox{helper} is the only child of \doublebox{helper},
\doublebox{helper} also becomes completed (\textcolor{blue}{\ding{172}}).
Then, since \doublebox{helper} is pointed to by three parental OR-nodes 
\singlebox{rev\_eq}, \singlebox{subg2}, and, \singlebox{repleh}), all of them become completed (\textcolor{blue}{\ding{173}}).

While we could stop the status update since the root node has been completed,
let us examine how the update propagates from other newly updated OR-nodes 
(\singlebox{subg2} and \singlebox{repleh}).
\singlebox{subg2} is pointed to by an OR-node (\doublebox{subg1 $\land$ subg2}),
but the other subgoal of \doublebox{subg1 $\land$ subg2} is not completed yet.
The dashed arrow labeled with \textcolor{blue}{\ding{174}} fails to complete
\doublebox{subg1 $\land$ subg2}, since \singlebox{subg1} is uncompleted.
On the other hand, the completion of \singlebox{repleh} completes its only parent AND-node 
(\doublebox{repleh}), 
since \singlebox{repleh} is the only child of \doublebox{repleh}, as depicted with the blue arrow labeled with \textcolor{blue}{\ding{174}}.
However, the upward propagation stops here, 
since both parents of \doublebox{repleh} are already completed either by the atomic proof for \singlebox{helper} or by the propagation labeled with \textcolor{blue}{\ding{173}} for
\singlebox{subg2}.

Since we completed the root node, we have a minimal subgraph sufficient to prove the root node, which is highlighted in Fig. \ref{fig:solution_graph_helper}.
Notice that even though we completed \singlebox{repleh}, \doublebox{repleh}, and \singlebox{subg2}, 
these nodes are not necessary to construct a proof of \singlebox{revs\_eq} and are not highlighted in Fig. \ref{fig:solution_graph_helper} for this reason.

\paragraph{Status Updates After Proving \texttt{repleh}.}

Fig. \ref{fig:update_cycle_repleh} on the other hand shows the case where we find an atomic proof for \singlebox{repleh}.
The atomic proof for \singlebox{repleh} makes this leaf OR-node completed.
Since \singlebox{repleh} is the only child of \doublebox{repleh},
\doublebox{repleh} also becomes completed (\textcolor{blue}{\ding{172}}).
Then, since \doublebox{repleh} is pointed to by two parental OR-nodes 
(\singlebox{subg2} and \singlebox{helper}), both of them become completed (\textcolor{blue}{\ding{173}}).

\singlebox{subg2} is pointed to by an OR-node (\doublebox{subg1 $\land$ subg2}),
but the other subgoal of \doublebox{subg1 $\land$ subg2} is uncompleted.
So, the dashed arrow labeled with \textcolor{blue}{\ding{174}} fails to complete \doublebox{subg1 $\land$ subg2}).

On the other hand, \singlebox{helper} is the only child of \doublebox{helper}.
So, the completion of \singlebox{helper} completes \doublebox{helper} (\textcolor{blue}{\ding{174}}).
\doublebox{helper} itself is pointed to by two AND-nodes (\singlebox{subg2} and \singlebox{revs\_eq}).
While the upward propagation stops at \singlebox{subg2} towards this direction,
since \singlebox{subg2} is already completed by \textcolor{blue}{\ding{173}} from \doublebox{repleh},
the completion of \doublebox{helper} completes the other parent node (\singlebox{revs\_eq}) through \textcolor{blue}{\ding{175}}.

Since \singlebox{revs\_eq} is the root node, we now have a minimal subgraph sufficient to prove the root node, which is highlighted in Fig. \ref{fig:solution_graph_repleh}.
Notice that even though we completed \singlebox{subg2}, this node is not necessary to construct a proof of \singlebox{revs\_eq} and is not highlighted in Fig. \ref{fig:solution_graph_repleh}.

\paragraph{Acyclic Solution Graphs}

These examples have demonstrably answered the question raised in Section \ref{sec:andor-graph}: \textit{Must our AND-OR graphs be acyclic, or can they be cyclic}?
Yes, abduction graphs can be cyclic to represent mutual dependencies among intermediate lemmas.
However, when completing the root node, the abduction graph must contain a subgraph sufficient to prove the root node, and that subgraph must be acyclic to avoid circular arguments.
We refer to such a minimal acyclic subgraph as a \textit{solution graph}.

\begin{figure}[t]
\begin{minipage}{0.48\textwidth}
\centering
\begin{tikzpicture}
[>={Stealth[round, scale=1.5]}, 
node distance=2cm, 
every edge/.style={draw, ->, text=blue}, 
every edge quotes/.style = {draw=none},
every node/.style={draw, ->, rounded corners, align=center}]
  \node[rounded corners] (A) {\verb|revs_eq|};
  \node[rounded corners, double, below=0.5cm of A] (C)
    { \texttt{subg1} $\land$ \texttt{subg2}};
  \node[below=0.5cm of C, xshift=-1.0cm] (C1) {\texttt{subg1}};
  \node[below=0.5cm of C, xshift=1.0cm] (C2) {\texttt{subg2}};
  \node[double, below=0.5cm of C2, xshift=-1.0cm] (C21) {\texttt{repleh}};
  \node[below=0.5cm of C21, fill=blue!25] (C211) {\texttt{repleh}};
  \node[double, below=0.5cm of C2, xshift=1.0cm] (C22) {\texttt{helper}};
  \node[below=0.5cm of C22] (C221){\texttt{helper}};
  \node[draw=none, below=-0.1cm of C211] (C2111) {\checkmark};
  \path
    (A) edge[] (C)
    (C) edge[] (C1)
    (C) edge[] (C2)
    (C2) edge[] (C21)
    (C2) edge[] (C22)
    (C21) edge[] (C211)
    (C22) edge[] (C221)
    (C211) edge[] (C22)
    (C21) edge[draw=blue, bend right=20] node[pos=0.75, below, draw=none] {\ding{173}} (C221)
    (C221) edge[] (C21)
    (C221) edge[draw=blue, bend right=40] node[pos=0.5, right, draw=none] {\ding{174}} (C22)
    (C2) edge[draw=blue, bend left=40, dashed] node[pos=0.5, left, draw=none] {\ding{174}} (C)
    (C21) edge[draw=blue, bend left=40] node[pos=0.5, left, draw=none] {\ding{173}} (C2)
    (C211) edge[draw=blue, bend left=40] node[pos=0.5, left, draw=none] {\ding{172}} (C21)
    (A) edge[bend left=35] (C22)
    (C22) edge[bend right=60, draw=blue] node[pos=0.5, right, draw=none] {\ding{175}} (A)
    ;
\end{tikzpicture}
\caption{\texttt{repleh} newly proved.}
\label{fig:update_cycle_repleh}
\end{minipage}
\hfill
\begin{minipage}{0.48\textwidth}
\centering
\begin{tikzpicture}
[>={Stealth[round, scale=1.5]}, 
node distance=2cm, 
every edge/.style={draw, ->}, 
every edge quotes/.style = {draw=none},
every node/.style={draw, ->, rounded corners, align=center}]
  \node[fill=blue!25] (A) {\verb|revs_eq|};
  \node[draw=gray, dashed, text=gray, double, below=0.5cm of A] (C)
    { \texttt{subg1} $\land$ \texttt{subg2}};
  \node[draw=gray, dashed, text=gray, below=0.5cm of C, xshift=-1.0cm] (C1) {\texttt{subg1}};
  \node[draw=gray, dashed, text=gray, below=0.5cm of C, xshift=1.0cm] (C2) { \texttt{subg2} };
  \node[double, below=0.5cm of C2, xshift=-1.0cm, fill=blue!25] (C21) { \texttt{repleh} };
  \node[below=0.5cm of C21, fill=blue!25] (C211) {\texttt{repleh}};
  \node[double, below=0.5cm of C2, xshift=1.0cm, fill=blue!25] (C22) { \texttt{helper} };
  \node[below=0.5cm of C22, fill=blue!25] (C221){\texttt{helper}};
  \node[draw=none, below=-0.1cm of C211] (C2111) {\checkmark};
  \path
    (A) edge[draw=gray, dashed] (C)
    (C) edge[draw=gray, dashed] (C1)
    (C) edge[draw=gray, dashed] (C2)
    (C2) edge[draw=gray, dashed] (C21)
    (C2) edge[draw=gray, dashed] (C22)
    (C21) edge[] (C211)
    (C22) edge[] (C221)
    (C211) edge[draw=gray, dashed] (C22)
    (C221) edge[] (C21)
    (A) edge[bend left=35] (C22)
    ;
\end{tikzpicture}
\Description{Solution graph in Fig. \ref{fig:update_cycle_repleh}.}
\caption{Solution graph in Fig. \ref{fig:update_cycle_repleh}.}
\label{fig:solution_graph_repleh}
\end{minipage}
\end{figure}

\section{System Description A: Recursive Abduction as the Expansion of Abduction Graph}\label{sec:recursive-abduction}

\begin{figure}[t]
\centering
\begin{minipage}{0.48\textwidth}
\begin{algorithm}[H]
\Description{\textsf{recursive abduction}}
\caption{\textsf{recursive abduction}}
\begin{algorithmic}[1] 
\STATE $graph \gets$ \textsf{set\_root} $goal$
\STATE $ctr \gets 1$
\WHILE{$ctr$ $\le$ limit $\land~\neg$proved $graph$}
\STATE $ctr \gets ctr + 1$
\STATE $nodes$ $\gets$ \textsf{get-contributive-leaves} $graph$
\STATE \textbf{fold} \textsf{work-on-node} $nodes$ $graph$ 
\ENDWHILE
\STATE show $graph$
\end{algorithmic}
\label{alg:recursive_abduction}
\end{algorithm}
\vspace{-20pt}
\end{minipage}
\begin{minipage}{0.48\textwidth}
\begin{algorithm}[H]
\caption{\textsf{work-on-node}}
\begin{algorithmic}[1]
\STATE $node$ $\gets$ \textsf{try-to-prove} $node$
\IF{\textsf{is-not-proved} $node$}
\STATE $graph$ $\gets$ \textsf{exp-tactic} $node$ $graph$
\STATE $graph$ $\gets$ \textsf{exp-template} $node$ $graph$
\STATE $graph$ $\gets$ \textsf{exp-mutation} $node$ $graph$
\ENDIF
\STATE $graph$ $\gets$ \textsf{update-is-completed} $node$ $graph$
\STATE $graph$ $\gets$ \textsf{update-is-contributive} $node$
\STATE \textbf{return} \textit{graph}
\end{algorithmic}
\label{alg:work-on-node}
\end{algorithm}
\vspace{-20pt}
\end{minipage}
\end{figure}

\begin{figure}[t]
\centering
\begin{minipage}{0.48\textwidth}
\begin{algorithm}[H]
\Description{\textsf{try-to-prove}}
\caption{\textsf{try-to-prove}}
\begin{algorithmic}[1] 
\STATE \hspace{0em}\textsf{Ors [}
\STATE \hspace{1em}\textsf{Thens [ Auto, IsSolved ],}
\STATE \hspace{1em}\textsf{Thens [}
\STATE \hspace{2em}\textsf{SmartInduct,}
\STATE \hspace{2em}\textsf{Ors [}
\STATE \hspace{3em}\textsf{Thens [ SimpAll, IsSolved ],}
\STATE \hspace{3em}\textsf{Thens [ Auto, IsSolved ]}
\STATE \hspace{2em}\textsf{]}
\STATE \hspace{1em}\textsf{]}
\STATE \hspace{1em}\textsf{Thens [ Hammer, IsSolved ]}
\STATE \hspace{0em}\textsf{]}
\end{algorithmic}
\label{alg:psl-try-to-prove}
\end{algorithm}
\vspace{-20pt}
\end{minipage}
\begin{minipage}{0.48\textwidth}
\begin{algorithm}[H]
\Description{\textsf{psl-exp-tactic} expansion}
\caption{\textsf{exp-tactic}}
\label{alg:exp-tactic}
\begin{algorithmic}[1] 
\STATE \hspace{0em}\textsf{Alts [}
\STATE \hspace{1em}\textsf{Clarsimp,}
\STATE \hspace{1em}\textsf{Thens [}
\STATE \hspace{2em}\textsf{SmartInduct,}
\STATE \hspace{2em}\textsf{Alts [ SimpAll, Auto ]}
\STATE \hspace{1em}\textsf{]}
\STATE \hspace{0em}\textsf{]}
\end{algorithmic}
\end{algorithm}
\vspace{-20pt}
\end{minipage}
\end{figure}

Section \ref{sec:abduction-graph} introduced the reasoning process over abduction graphs.
Now, we present the algorithm for building abduction graphs to search for proof scripts.
First, we describe the general workflow as Algorithm \ref{alg:recursive_abduction} and \ref{alg:work-on-node}, deferring the concrete definitions of certain functions to Section \ref{sec:try-to-prove} and \ref{sec:exp_tac}.

\subsection{The Main Loop for Recursive Abduction.}\label{sec:main-loop}

The AbductionProver's overall flow is characterized by its ability to expand the abduction graph recursively by producing contributive conjectures.
This recursive nature is realized by the \texttt{while} rule in Algorithm \ref{alg:recursive_abduction}.

Algorithm \ref{alg:recursive_abduction} first initializes the graph with the root OR-node, representing the final goal provided by users, in Line 1. Then, it sets the counter (\textit{ctr}) to 1.

In the main \texttt{while} loop, Algorithm \ref{alg:recursive_abduction} increments the counter (Line 4) and fetches a subset of OR-leaf nodes that are still needed to prove the root node (Line 5).
Then, it applies \texttt{work-on-node} to for each such OR-node (Line 6).

In Algorithm \ref{alg:work-on-node}, \texttt{work-on-node} first attempts to find an atomic proof for the given OR-leaf node, using \texttt{try-to-prove}, which we discuss in Section \ref{sec:try-to-prove} (Line 1).
Then, if \texttt{try-to-prove} fails to 
find an atomic proof for the OR-leaf node, Algorithm \ref{alg:work-on-node} expands the abduction graph for this node 
by applying tactics (Line 3), generating conjectures using predefined templates (Line 4), or employing mutation algorithms (Line 5).

Regardless of whether \texttt{work-on-node} successfully proves the OR-node or expands the graph for it, the algorithm updates the graph based on the following information: the proof status of the OR-node and the addition of new edges and nodes from one of the \texttt{exp-} functions. This update is performed using the procedures discussed in Section \ref{sec:completion} and Section \ref{sec:contributive} (Lines 7–8). 
Algorithm \ref{alg:recursive_abduction} exits the main \texttt{while} loop when the counter reaches a certain threshold or when the root node is completed.

Now we look more closely to the concrete functions, \textsf{try-to-prove}, \textsf{exp-tactic}, \textsf{exp-template}, and \textsf{exp-mutation}. 
Notably, the first two functions internally utilize the OR-tree model introduced in Section \ref{sec:ortree-tactic}. 
Since their primary focus is on specific OR-nodes rather than the dependencies within the overall abduction graph, we leverage the simplicity of the OR-tree model to employ the PSL \cite{psl} framework for small-scale automation.

\subsection{\textsf{try-to-prove}: Attempt to Find Atomic Proofs For OR-leaf Nodes.}\label{sec:try-to-prove}

To build a solution graph within an abduction graph,
one has to find atomic proofs for certain OR-leaf nodes. 
To achieve this, we employ a proof strategy in PSL \cite{psl}, as outlined in Algorithm \ref{alg:psl-try-to-prove}. 

\texttt{try-to-prove} implements a backtracking search over three sub-strategies:
one based on  a general-purpose tactic (\verb|auto|), 
the second based on proof by induction, and the last one using \verb|sledgehammer| \cite{sledgehammer}.
Its search path of \texttt{try-to-prove} is shown in Fig. \ref{fig:try-to-prove} schematically.
In this figure, numbers in the edge labels represent the order of implicit tree traversal, and dashed edges represent the failures of tactic applications, which triggers backtracking.
As the numbers on the edges and the single-lined node edges illustrate,
the PSL's interpreter executes the depth-first search based on \texttt{try-to-prove} but fails to find a proof for \texttt{revs\_eq}.

The \texttt{Ors} combinator implements deterministic choice, and as soon as the PSL interpreter finds an atomic proof using Algorithm \ref{alg:psl-try-to-prove}, it stops the search, as a single atomic proof suffices to complete the OR-node.

\begin{figure}[t]
\centering
\begin{tikzpicture}
[>={Stealth[round, scale=1.0]}, 
node distance=2cm, 
every edge/.style={draw, ->}, 
every edge quotes/.style = {auto, draw=none},
every node/.style={draw, ->}]
  \node[rounded corners] (A) {\texttt{revs\_eq. rev1 $xs$ [] = rev2 $xs$}};
  \node[rounded corners, below=1.5cm of A, xshift=-6.0cm, align=left, dashed] (B) {\texttt{error}};
  \node[rounded corners, below=1.5cm of A, xshift=-2.5cm, align=left] (C) {\texttt{subg1'$\land$subg2'}};
  \node[rounded corners, below=1.5cm of A, xshift=2.5cm, align=left] (D) {\texttt{subg1$\land$subg2}};
  \node[rounded corners, below=1.5cm of A, xshift=6.0cm, align=left, dashed] (E) {\texttt{error}};
  \node[rounded corners, below=1.5cm of C, xshift=-3.0cm, align=left] (C1) {\texttt{subgoals}};
  \node[rounded corners, below=1.5cm of C, xshift=0.0cm, align=left] (C2) {\texttt{subgoals}};
  \node[rounded corners, below=1.5cm of D, xshift=-2.5cm, align=left] (D1) {\texttt{subgoals}};
  \node[rounded corners, below=1.5cm of D, xshift=1.5cm, align=left] (D2) {\texttt{subgoals}};
  \node[rounded corners, below=1.0cm of C1, align=left, dashed] (C11) {\texttt{error}};
  \node[rounded corners, below=1.0cm of C2, align=left, dashed] (C21) {\texttt{error}};
  \node[rounded corners, below=1.0cm of D1, align=left, dashed] (D11) {\texttt{error}};
  \node[rounded corners, below=1.0cm of D2, align=left, dashed] (D21) {\texttt{error}};
  \path
    (A) edge[dashed, sloped, "1. \texttt{auto}"] (B)
    (A) edge[pos=0.6, sloped, "2. \texttt{induct}"] (C)
    (A) edge[pos=0.65, sloped, "7. \texttt{induct $xs$}"] (D)
    (A) edge[pos=0.7, dashed, sloped, "12. \texttt{sledgehammer}"] (E)
    (C) edge[pos=0.7, sloped, "3. \texttt{simp\_all}"] (C1)
    (C) edge[sloped, "5. \texttt{auto}"] (C2)
    (D) edge[pos=0.7, sloped, "8. \texttt{simp\_all}"] (D1)
    (D) edge[sloped, "10. \texttt{auto}"] (D2)
    (C1) edge[dashed, auto, "4. \texttt{is\_solved}"] (C11)
    (C2) edge[dashed, auto, "6. \texttt{is\_solved}"] (C21)
    (D1) edge[dashed, auto, "9. \texttt{is\_solved}"] (D11)
    (D2) edge[dashed, auto, "11. \texttt{is\_solved}"] (D21)
    ;
\end{tikzpicture}
\Description{A potential search tree of \texttt{try-to-prove}.}
\caption{The implicit OR-tree exploration of \texttt{try-to-prove} for \texttt{revs\_eq}.}
\label{fig:try-to-prove}
\end{figure}

\subsection{Implicit Conjecturing using \textsf{exp-tactic}: Tactics to Expand OR-leaf nodes.}\label{sec:exp_tac}

Contrary to \texttt{try-to-prove} in Algorithm \ref{alg:psl-try-to-prove}, \texttt{exp-tactic} in Algorithm \ref{alg:exp-tactic} attempts to expand the OR-node by taking all successful leaf nodes as shown in Fig. \ref{fig:exp-tactic-to-ortree}.
This time, only one leaf is surrounded by dashed boundary and pointed to by a dashed edge.
That is, 
only one leaf node represents the complete failure of tactic application,
while other nodes contain viable subgoals that are at least aligned with
MP-tac from Section \ref{sec:ortree-tactic}.
That is, even though these leaves do not represent the completion of proof search either; they can be used to expand the abduction graph.

\begin{figure}[t]
\centering
\begin{tikzpicture}
[>={Stealth[round, scale=1.0]}, 
node distance=2cm, 
every edge/.style={draw, ->, sloped}, 
every edge quotes/.style = {auto, draw=none},
every node/.style={draw, ->, rounded corners, align=left}]
  \node[] (A) {\texttt{revs\_eq. rev1 $xs$ [] = rev2 $xs$}};
  \node[below=1.5cm of A, xshift=-6.0cm, dashed] (B) {\texttt{error}};
  \node[below=1.5cm of A, xshift=-2.5cm] (C) {\texttt{subg1'$\land$subg2'}};
  \node[below=1.5cm of A, xshift=3.5cm] (D) {\texttt{subg1$\land$subg2}};
  \node[below=1.5cm of C, xshift=-3.0cm] (C1) {\texttt{subgoals}};
  \node[below=1.5cm of C, xshift=0.0cm] (C2) {\texttt{subgoals}};
  \node[below=1.5cm of D, xshift=-2.5cm, fill=yellow!50] (D1) {\texttt{subgoals}};
  \node[below=1.5cm of D, xshift=1.5cm] (D2) {\texttt{subgoals}};
  \path
    (A) edge[dashed, "1. \texttt{clarsimp}"] (B)
    (A) edge[pos=0.6, "2. \texttt{induct}"] (C)
    (A) edge[pos=0.65, "5. \texttt{induct $xs$}"] (D)
    (C) edge[pos=0.7, "3. \texttt{simp\_all}"] (C1)
    (C) edge["4. \texttt{auto}"] (C2)
    (D) edge[pos=0.7, "6. \texttt{simp\_all}"] (D1)
    (D) edge["7. \texttt{auto}"] (D2)
    ;
\end{tikzpicture}
\Description{Expansion of \texttt{revs\_eq} using \texttt{exp-prove}.}
\caption{Expansion of \texttt{revs\_eq} using Algorithm \ref{alg:exp-tactic} (\texttt{exp-tactic}).}
\label{fig:exp-tactic-to-ortree}
\end{figure}

Note that both Algorithm \ref{alg:psl-try-to-prove} and Algorithm \ref{alg:exp-tactic} are executed by PSL, which is based on the OR-tree model.
As such, each node produced by Algorithm \ref{alg:exp-tactic} may or may not contain multiple subgoals.
For this reason, the OR-nodes generated by Algorithm \ref{alg:exp-tactic} are first incorporated into the abduction graph as AND-nodes.
Then, the Prover decomposes each AND-node from Algorithm \ref{alg:exp-tactic} into separate OR-nodes within the abduction graph, after confirming that treating the remaining subgoals as independent conjectures allows us to complete the newly added AND-node.

This transition from the local use of the OR-tree model to the global use of the AND-OR graph model is illustrated in Fig. \ref{fig:exp-tactic-to-andortree}.
This figure integrates the OR-leaf node, highlighted with a yellow background in Fig. \ref{fig:exp-tactic-to-ortree}, as a yellow AND-node in Fig. \ref{fig:exp-tactic-to-andortree}.
Note that the label from \singlebox{revs\_eq} to \doublebox{subgoals} in Fig. \ref{fig:exp-tactic-to-andortree} is the concatenation of the labels from \singlebox{revs\_eq} to \singlebox{subgoals} in Fig. \ref{fig:exp-tactic-to-ortree}, followed by the tactic application (\texttt{simp add: subgoal}) required to complete \doublebox{subgoals}, assuming \singlebox{subgoals}.

Notice that it is not guaranteed that the \texttt{simp} tactic can always accomplish this task in the general setting.
Therefore, the AbductionProver determines the necessary tactics and their arguments by running \texttt{sledgehammer} on the AND-node after temporarily assuming all its child OR-nodes in the background context using the PSL strategy called \texttt{jackhammer} described in Algorithm \ref{alg:jackhammer}.

This PSL strategy invokes \texttt{sledgehammer} as many times as possible until \texttt{sledgehammer} fails.
For this specific subgoal, it calls \texttt{sledgehammer} once, obtains the tactic (\texttt{simp add: subgoals}), and stops because this tactic application discharges the only remaining subgoal.


\begin{figure}[t]
\centering
\begin{tikzpicture}
[>={Stealth[round, scale=1.0]}, 
node distance=2cm, 
every edge/.style={draw, ->}, 
every edge quotes/.style = {draw=none},
every node/.style={draw, ->, rounded corners}]
  \node[] (N1) {\texttt{revs\_eq. rev1 $xs$ [] = rev2 $xs$}};
  \node[double,  below=1.0cm of N1, fill=yellow!50] (N2) {\texttt{subgoals}};
  \node[below=0.5cm of N2] (N3) {\texttt{subgoals}};
  \node[draw=none, below=0.1cm of N1, xshift=-4.0cm] (L) {\dots};
  \node[draw=none, below=0.1cm of N1, xshift=4.0cm] (R) {\dots};
  \path
    (N1) edge[midway, "\texttt{induct $xs$, simp\_all, simp add: subgoal}"] (N2)
    (N2) edge[] (N3)
    (N1) edge[] (L)
    (N1) edge[] (R)
    ;
\end{tikzpicture}
\Description{Integration of OR-nodes into Abduction Graph.}
\caption{Integration of OR-nodes into Abduction Graph.}
\label{fig:exp-tactic-to-andortree}
\end{figure}
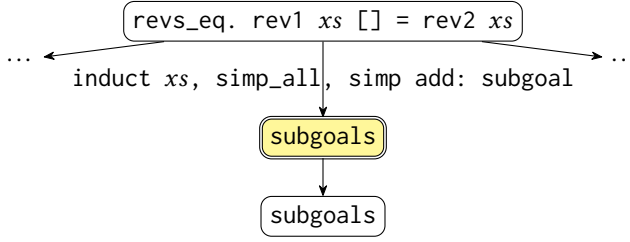

\begin{figure}[t]
\centering
\begin{minipage}{0.48\textwidth}
\begin{algorithm}[H]
\Description{\textsf{Temporarily Prove AND-nodes by Assuming OR-nodes.}}
\caption{\texttt{jackhammer}} 
\begin{algorithmic}[1] 
\STATE \hspace{0em}\textsf{Thens [}
\STATE \hspace{1em}\textsf{Repeat ( Hammer ),}
\STATE \hspace{1em}\textsf{IsSolved}
\STATE \hspace{0em}\textsf{]}
\end{algorithmic}
\label{alg:jackhammer}
\end{algorithm}
\vspace{-20pt}
\end{minipage}
\begin{minipage}{0.48\textwidth}
\begin{algorithm}[H]
\Description{quickpick}
\caption{\texttt{refute}}
\label{alg:quickpick}
\begin{algorithmic}[1] 
\STATE \hspace{0em}\textsf{Thens [}
\STATE \hspace{1em}\textsf{Quickcheck,}
\STATE \hspace{1em}\textsf{Nitpick}
\STATE \hspace{0em}\textsf{]}
\end{algorithmic}
\end{algorithm}
\vspace{-20pt}
\end{minipage}
\end{figure}

\section{System Description B: Explicit Conjecturing}\label{sec:explicit-conjecturing}

Section \ref{sec:recursive-abduction} introduced how to gradually expand abduction graphs using PSL and tactics.
As discussed in Section \ref{sec:tactic_as_conjecturing}, we consider such tactic applications as a form of \textit{implicit} conjecturing.

The key distinction between implicit and explicit conjecturing lies in how the relevance of conjectures (denoted as \textsf{C1} in Section \ref{sec:one-step}) is verified.
Implicit conjecturing relies on the underlying prover’s mechanism to ensure that the left-hand branch of the derivation tree (MP-tac) in Section \ref{sec:ortree-tactic} holds, thereby automatically granting the relevance of conjectures.
In contrast, explicit conjecturing requires us to explicitly filter conjectures based on their relevance by proving the left-hand branch of the derivation tree (MP-conj) in Section \ref{sec:one-step}, as we will see in Section \ref{sec:bunched-conjeturing}.

Currently, the AbductionProver employs two approaches for \textit{explicit} conjecturing: template-based conjecturing and mutation-based conjecturing.

\subsection{\texttt{exp-template}: Template-Based Conjecturing.}

In template-based conjecturing, embodied as \texttt{exp-template}, AbductionProver first collects relevant functions and generates conjectures based on 16 pre-defined templates. 
These templates parametrically describe common patterns such as associativity, commutativity, reflexivity, and distributivity, without being specific to any particular problem domain.

While our template-based conjecturing approach in AbductionProver builds upon the earlier work of Nagashima et al., the direction of reasoning differs fundamentally. Nagashima et al. employed template-based conjecturing in a bottom-up manner, focusing on generating and proving conjectures from available functions and data types using the default, extensive PSL strategy, independently of the original goal.

In contrast, AbductionProver adopts a top-down goal-oriented approach, assessing the relevance of conjectures via Modus Ponens, as discussed in Section \ref{sec:one-step}. Even if the default, less extensive PSL strategy (Algorithm \ref{alg:psl-try-to-prove}) fails to establish an atomic proof for a conjecture, AbductionProver retains it as an OR-leaf, enabling further proof efforts through recursive conjecturing.

A known limitation of the template-based approach discussed in the literature is its lack of specificity to a given problem.
For example, the \texttt{helper} lemma from Section \ref{sec:one-step} does not fit into any of the predefined templates, yet it closely resembles the original goal, \texttt{revs\_eq}.

\subsection{\texttt{exp-mutation}: Mutation-Based Conjecturing.}

Mutation-based conjecturing addresses this limitation by generating conjectures through mutations of the current goals.
Since all mutants originate from the goals (OR-leaves), this approach aims to maintain a degree of specificity to them. 
Currently, the AbductionProver employs the following six mutation algorithms:

\paragraph{1: Remove outermost assumption.} 
Strips away the outermost implication (or assumption) to obtain a more direct statement. This may remove useful terms from the assumption that could help prove the conclusion, but it may also provide a clearer focus on the conclusion for the subsequent steps. 
Example:
\begin{alltt}
Input:  rev1 \(xs\) [] = rev2 \(xs\) \(\Longrightarrow\) rev1 (\(x1\):\(xs\)) [] = rev2 (\(x1\):\(xs\))
Output: rev1 (\(x1\):\(xs\)) [] = rev2 (\(x1\):\(xs\))
\end{alltt}

\paragraph{2: Remove function}
Strips away the outermost implication (or assumption) to obtain a more direct statement. This may remove useful terms from the assumption that could help prove the conclusion, but it may also provide a clearer focus on the conclusion for the subsequent steps. Example:

\begin{alltt}
Input:  rev1 \(xs\) \(ys\) = rev2 \(xs\) + \(ys\)
Output: rev1 \(xs\) \(ys\) = \(xs\) + \(ys\)
\end{alltt}

\paragraph{3: Abstract same term.}
Identifies identical or repeated sub-terms and replaces them with a fresh variable.
This reduces redundancy and highlights the common sub-term, making the goal potentially simpler to manipulate. However, it may also eliminate information that is useful for completing the proof.
Example:

\begin{alltt}
Input:  rev1 \(xs\) [] = rev2 \(xs\) \(\Longrightarrow\) rev1 (\(x1\):\(xs\)) [] = rev2 (\(x1\):\(xs\))
Output: rev1 \(xs\) [] = rev2 \(xs\) \(\Longrightarrow\) rev1 \(ys\) [] = rev2 \(ys\)
\end{alltt}

\paragraph{4: Replace implication with equation.}
Rewrites an implication as an equation. 
In general, this makes the goal harder to prove; however, the stronger goal based on an equation can occasionally open up alternative inference paths or simplifications.
Example:
\begin{alltt}
Input:   rev1 \(xs\) [] = rev2 \(xs\) \(\Longrightarrow\) rev1 (\(x1\):\(xs\)) [] = rev2 (\(x1\):\(xs\))
Output: (rev1 \(xs\) [] = rev2 \(xs\)) = (rev1 (\(x1\):\(xs\)) [] = rev2 (\(x1\):\(xs\)))
\end{alltt}

\paragraph{5: Generalize by renaming variables.}
Renames some occurrences of variables to fresh variables when the goal contains multiple occurrences of the same variable.
While this mutation makes the goal stronger, the resulting statement is sometimes better aligned with certain proof procedures, such as proof by induction.

\begin{alltt}
Input:   rev1 \(xs\) [] = rev2 \(xs\) \(\Longrightarrow\) rev1 (\(x1\):\(xs\)) [] = rev2 (\(x1\):\(xs\))
Output:  rev1 \(xs\) [] = rev2 \(ys\) \(\Longrightarrow\) rev1 (\(x1\):\(xs\)) [] = rev2 (\(x1\):\(ys\))
\end{alltt}

\paragraph{6: Generalize, then extend.}
Generalizes the goal by replacing constants with variables,
then extends it by replacing a subterm of the goal with a composite term
that applies a function to the original subterm.
This mutation algorithm selects such functions from the list of registered functions in the proof context
based on their relevance to the functions appearing in the input goal.
Currently, the algorithm considers functions that appear in the definitions of the used functions as relevant.
For the following example, the algorithm selects the function (\texttt{+}) from the proof context
because this function is used in the definition of \texttt{rev2}.
Example:

\begin{alltt}
Input:       rev1 \(xs\) [] = rev2 \(xs\)
Generalized: rev1 \(xs\) \(ys\) = rev2 \(xs\)
Output:      rev1 \(xs\) \(ys\) = rev2 \(xs\) + \(ys\)
\end{alltt}
\noindent
This mutation was first proposed by Nagashima \etal{} \cite{pgt} and sometimes produces broader statements that relate previously unconnected parts of the proof context.

Note that although we have shown only one example output for each input,
these algorithms generally produce numerous candidate conjectures from a single subgoal.

\noindent

\section{System Description C: Conjecture Filtering}\label{sec:filtering}

The aforementioned conjecturing approaches, particularly the mutation-based algorithm, tend to generate a large number of candidates.
While the graph structure presented in Section \ref{sec:background} helps mitigate the combinatorial explosion by enabling node-sharing,
the sheer volume of candidates still necessitates additional filtering strategies.
As such, the AbductionProver employs many filtering approaches step-by-step.

\subsection{Abductive Reasoning: Filtering Based on Modus Ponens}

The central filtering method of the AbductionProver is based on the Modus Ponens as discussed in Section \ref{sec:one-step}:
we want conjectures that are useful (\texttt{C1}) but not equivalent to \texttt{False} (\textsf{C2}).

\paragraph{Filtering based on Counterexample Finders.}

Similarly to previous work in this field, we use available counterexample finders (\texttt{Quickcheck} and \texttt{Nitpick}) to filter out false conjectures using the PSL program, \texttt{refute}, shown in Algorithm \ref{alg:quickpick}.
Algorithm \ref{alg:quickpick} sequentially combines two sub-strategies: \texttt{Quickcheck} and \texttt{Nitpick}.
This ordering ensures that the Prover invokes \texttt{Nitpick} only if \texttt{Quickcheck} finds no counterexamples, thereby improving overall performance, as \texttt{Nitpick} is often more computationally expensive than \texttt{Quickcheck}.

The results are stored in a lookup table, separate from the main data structure—the abduction graph introduced in Section \ref{sec:abduction-graph}, to avoid repeatedly invoking the resource, intensive counterexample finders on identical conjectures generated from different branches of the abduction graph.

This use of counterexample finders addresses the right branch of the derivation tree (MP-conj) in Section \ref{sec:one-step}, which corresponds to the \textsf{C2} requirement presented in the same section.

\paragraph{One-Step Abduction.}
On the other hand, one-step abduction addresses the left branch of the derivation tree (MP-conj) in Section \ref{sec:one-step}, 
which corresponds to the \textsf{C1} requirement presented in the same section.
The AbductionProver achieves this by temporarily assuming the conjectures under consideration and attempting to prove the goal using these assumptions.

Contrary to previous work on abductive reasoning in Isabelle \cite{pgt}, which inserts conjectures as assumptions within the corresponding goals and attempts to prove the resulting goals using the \texttt{fastforce} tactic, the AbductionProver instead registers conjectures axiomatically in the underlying proof context and attempts to prove the respective goal within this context using \texttt{sledgehammer}.

This shift enables the AbductionProver not only to leverage the strong proof automation of \texttt{sledgehammer} but also to efficiently exploit multiple conjectures simultaneously, identifying useful conjecture combinations rather than a single useful conjecture, using successive proof-guided conjecture set refinement explained below.

\subsection{Successive Proof-Guided Conjecture Set Identification using \texttt{sledgehammer}}\label{sec:bunched-conjeturing}

When we introduced one-step conjecturing in Section \ref{sec:one-step}, it was presented as if the AbductionProver would check the relevance (\textsf{C1}) of each single candidate conjecture separately.
Indeed, that was the approach taken by the previous work on abductive reasoning in Isabelle \cite{pgt}; however, checking individual conjectures in isolation suffers from two key issues:

\begin{itemize}[noitemsep, topsep=0pt, leftmargin=*]
    \item \textit{Scalability concerns:} When numerous conjectures are present, evaluating the left branch of MP-conj (\texttt{conj $\Longrightarrow$ goal}) separately for each conjecture leads to significant performance degradation.
    \item \textit{Dependency among conjectures:} While some goals require a set of conjectures simultaneously to be completed, a naive application of explicit single conjecturing does not inherently provide these sets. 
\end{itemize}

The second limitation stands in stark contrast to tactic applications, which act as implicit conjecturing: a single tactic application may generate multiple subgoals (conjectures), all of which must be proven. In contrast, when generating multiple conjectures explicitly, it is not immediately clear which subsets must be proven. The number of potential subsets grows combinatorially, making brute-force evaluation impractical for large conjecture spaces.

\begin{figure}[t]
\begin{tikzpicture}
[>={Stealth[round, scale=1.0]}, 
node distance=2cm, 
every edge/.style={draw, ->}, 
every node/.style={draw, ->, rounded corners, align=center}]
\node[draw=none] (R1) {round 1};
\node[draw=none, right=1.5cm of R1] (R2) {round 2};
\node[draw=none, right=1.5cm of R2] (R3) {round 3};
\node[draw=none, right=1.5cm of R3] (R4) {round 4};
\node[below=0.3cm of R1] (N1) 
 {\begin{tabular}{c} 
   conjecture-1 \\ 
   \colorbox{lightpink}{conjecture-2} \\ 
   \colorbox{lightpink}{conjecture-3} \\ 
   conjecture-4 
  \end{tabular}};
\node[below=0.3cm of R2] (N11) 
 {\begin{tabular}{c} 
   conjecture-1 \\ 
   \colorbox{lightpink}{conjecture-2} \\ 
   conjecture-4 
  \end{tabular}};
\node[right=0.5cm of N1, below=0.1cm of N11] (N12) 
 {\begin{tabular}{c} 
   conjecture-1 \\ 
   \colorbox{lightpink}{conjecture-3} \\ 
   \colorbox{lightpink}{conjecture-4} 
  \end{tabular}};
\node[below=0.3cm of R3] (N111) 
 {\begin{tabular}{c} 
   conjecture-1 \\  
   \colorbox{lightpink}{conjecture-3}
  \end{tabular}};
\node[below=0.5cm of N111] (N112) 
 {\begin{tabular}{c} 
   conjecture-1 \\  
   conjecture-4 
  \end{tabular}};
\node[right=0.5cm of N111] (N1111) 
 {\begin{tabular}{c} 
   conjecture-1
  \end{tabular}};
\path
(N1) edge[] (N11)
(N1) edge[] (N12)
(N11) edge[] (N112)
(N12) edge[] (N111)
(N12) edge[] (N112)
(N111) edge[] (N1111)
;
\end{tikzpicture}
\Description{Successive Proof-Guided Conjecture Set Identification. In these four rounds, the AbductionProver identifies 4 sets of provably useful conjectures: \{conjecture-2\}, \{conjecture-2, conjecture-3\}, \{conjecture-3\}, \{conjecture-3, conjecture-4\}.}
\caption{Successive Proof-Guided Conjecture Set Identification. In these four rounds, the AbductionProver identifies 4 sets of provably useful conjectures: 
\{conjecture-2\}, \{conjecture-2, conjecture-3\}, \{conjecture-3\}, \{conjecture-3, conjecture-4\}.}
\label{fig:bunched-conjecturing}
\end{figure}
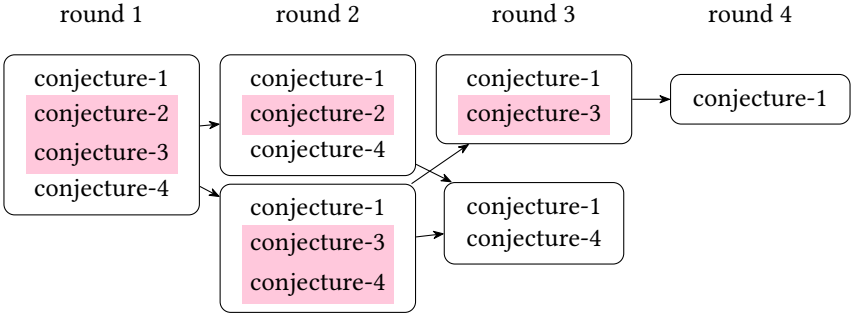

To address these issues, we developed a method called \textit{successive proof-guided conjecture set identification}.
In this approach, we temporarily assume all conjectures that pass the initial screening and register them in the underlying proof context.
We then run \texttt{sledgehammer} to determine whether it can prove the current goal using any of these assumed conjectures.

If \texttt{sledgehammer} finds a proof script, AbductionProver extracts the conjectures used in the proof and generates alternative conjecture sets by removing one used conjecture from the previous set for each alternative set.
This process repeats iteratively until either \texttt{sledgehammer} fails to find a proof for all candidate subsets or the iteration reaches a predefined limit, currently set to 15 for the root node and 5 for other OR-nodes

This iterative process is illustrated in Fig. \ref{fig:bunched-conjecturing} schematically using an abstract scenario.
This figure shows that the process starts with the four conjectures each named conjecture-$n$. 
In the first round, \texttt{sledgehammer} finds a proof script involving two of them (conjecture-2 and conjecture-3) and registers this subset as a candidate AND-node.
Then, it produces two subsets by removing one of these conjectures for each subset.

In the second round, the process runs \texttt{sledgehammer} with these subsets of cardinality 3 against the same goal. 
For the first subset, \texttt{sledgehammer} found a script involving conjecture-2, while it found a script involving conjecture-3 and conjecture-4.
Therefore, the process registers the two subsets ({conjecture-2} and {conjecture-3, conjecture-4}) as candidate for AND-nodes.
The third round and fourth round proceed similarly, while each round involves subsets using which \texttt{sledgehammer} fails to prove the current goal, leading to the end of the iterative process.

Note that despite its iterative nature, successive proof-guided conjecture set identification constitutes a single one-step conjecturing process.
For example, the four sets of conjectures identified in Fig. \ref{fig:bunched-conjecturing} serve as candidate child AND-nodes for a single parent OR-node.
Within each iteration of the main loop, presented in Section \ref{sec:main-loop} as Algorithm \ref{alg:recursive_abduction}, AbductionProver applies successive proof-guided conjecture set identification to each OR-node selected by \textsf{get-contributive-leaves}.

\subsection{Filtering Beyond Abductive Reasoning}\label{sec:beyond-abduction}

Successive proof-guided conjecture set identification generates many conjectures that are either false or redundant. To mitigate combinatorial explosion, the AbductionProver incorporates several additional filtering mechanisms beyond abductive reasoning, including:  type-guided conjecturing, $\alpha$-normalization, simplifier-normalization, and history-sensitive pruning. Due to space limitations, we refrain from describing these mechanisms in detail in this paper.

\section{Conclusion}\label{sec:conclusion}

This paper presented the AbductionProver for Isabelle/HOL, a proof-search framework based on recursive abductive reasoning over AND-OR graphs.
The framework uniformly treats tactic applications and explicit conjecturing as instances of Modus Ponens, enabling proof search through the expansion of abduction graphs while tracking completion and contributivity statuses of nodes.

We extended the standard OR-tree interpretation of tactical theorem proving to AND-OR graphs to support recursive conjecturing and the sharing of intermediate lemmas across proof branches.
The AbductionProver combines local OR-tree exploration using PSL with global graph-based reasoning, supports cyclic abduction graphs together with acyclic solution extraction, and integrates multiple techniques for conjecture generation, normalization, and pruning.

This paper focused on clarifying the computational model and architectural design of the AbductionProver through a running example and algorithmic descriptions.
A working implementation for Isabelle/HOL is publicly available.

Overall, the current implementation suggests that graph-based abductive reasoning is a promising direction for proof automation in tactical theorem provers.

\bibliographystyle{ACM-Reference-Format}
\bibliography{main}

\end{document}